\let\latexaddtocontents\addtocontents
\documentclass[a4paper,twocolumn,11pt, unpublished]{quantumarticle}
\let\addtocontents\latexaddtocontents
\pdfoutput=1

\usepackage{xcolor}
\definecolor{babyblue}{rgb}{0.54, 0.81, 0.94}

\usepackage[utf8]{inputenc}
\usepackage{csquotes}
\usepackage[english]{babel}
\usepackage[T1]{fontenc}

\usepackage{amsmath}
\usepackage{amssymb}
\usepackage{bm} 
\usepackage{braket}

\usepackage{graphicx} 
\usepackage{dcolumn} 
\usepackage{subcaption} 
\usepackage{tikz}
\usepackage{pgfplots}
\usepackage{lipsum}
\usepackage{titlesec} 
\usepackage{setspace} 
\pgfplotsset{compat=1.17}

\usepackage{hyperref}

\usepackage[style=numeric, sorting=none]{biblatex}
\begin{document}

\title{Near-Term Quantum Spin Simulation of the Spin-$\frac{1}{2}$ Square $J_{1}-J_{2}$ Heisenberg Model}

\author{Dylan Sheils}
\affiliation{Department of Physics, Applied Physics and Astronomy, Rensselear Polytechnic Institute, Troy, NY 12180}
\author{Trevor David Rhone}
\affiliation{Department of Physics, Applied Physics and Astronomy, Rensselear Polytechnic Institute, Troy, NY 12180}
\maketitle

\begin{abstract}
  
  Simulating complex spin systems, known for high frustration and entanglement, presents significant challenges due to their intricate energy landscapes. This study focuses on the $J_{1}-J_{2}$ Heisenberg model, renowned for its rich phase behavior on the square lattice, to investigate strongly correlated spin systems. We conducted the first experimental quantum computing study of this model using the 127-qubit IBM Rensselear Eagle processor and the Variational Quantum Eigensolver (VQE) algorithm. By employing classical warm-starting ($+40\%$ ground state energy approximation) and a newly developed ansatz ($+9.31\%$ improvement compared to prior best), we improved ground state approximation accuracy on the 16-site variant, achieving usable results with approximately $10^{3}$ iterations, significantly fewer than the $10^{4}-10^{5}$ steps proposed by previous theoretical studies. We utilized existing error mitigation strategies and introduced a novel Classically-Reinforced VQE error mitigation scheme, achieving $93\%$ ground state accuracy, compared to $83.7\%$ with the Quantum Moments algorithm and $60\%$ with standard error mitigation. These strategies reduced the average error of observable prediction from $\approx 20\%$ to $5\%$, enhancing phase prediction from qualitative to quantitative alignment. Additionally, we explored an experimental implementation of the Quantum Lanczos (QLanczos) algorithm using Variational-Fast Forwarding (VFF) on a 4-qubit site, achieving $\approx 97\%$ ground state approximation. Theoretical simulations indicated that Krylov-based methods outperform VQE, with the Lanczos basis converging faster than the real-time basis. Our study demonstrates the capability of near-term quantum devices to predict phase-relevant observables for the $J_1-J_2$ Heisenberg model, transitioning focus from theoretical to experimental, and suggesting general improvements to VQE-based methods.
\end{abstract}

Spin system simulation presents a unique computational challenge for classical methods. \cite{nakamura2019strategy, huerga2022variational,
harris2018phase,
isaev2009hierarchical, 
arlego2008plaquette, PhysRevResearch.3.033002,
10.1063/1.3518900} Applying theoretically exact methods tends to fail due to high computational cost, while classical approximate and classical variational methods tend to perform inconsistently. \cite{nakamura2019strategy, huerga2022variational,
harris2018phase,
isaev2009hierarchical, 
arlego2008plaquette} Approximate methods face several fundamental issues, which may prevent their accurate simulation of spin systems: 1) the methods can fail to capture the entanglement structure describing the dynamics of these systems due to the system's large dimensionality \cite{PhysRevResearch.3.033002} and 2) the methods may project to an non-optimal low-energy subspace thereby simulating an misrepresentative excited state, due to the exponentially decreasing spectral gap and high degeneracy \cite{10.1063/1.3518900, nakamura2019strategy}. To address these challenges, the current work aims to explore various quantum computing approaches to accurately simulate the square $J_{1}-J_{2}$ system, achieving phase predictions under experimentally feasible circuit depths and iteration numbers. By shifting the focus from theoretical performance predictions of future fault-tolerant systems to the practical performance of near-term systems, and by identifying the most promising simulation methods, we hope to guide future research towards realizing quantum advantage.

We will explore the Variational Quantum Eigensolver (VQE) and the Quantum Lanczos (QLanczos) algorithm. We will also present a novel Classically-Reinforced ZNE correction scheme. For QLanczos, the circuit depth is a function of the $k$ Lanczos vectors which, for large $k$ especially, becomes challenging for current quantum computers. To combat this, we employ Variational Fast-Forwarding (VFF) to approximate the necessary circuit, removing the $k$ Lanczos vector performance penalty. \cite{filip2024variational} The physical system being examined in this study is the square $J_{1}-J_{2}$ Heisenberg model, Eq.~\ref{ham}. 

\begin{align}
  \label{ham}
   H = J_{1}\sum_{i, j \in N(i)} S_{i} \cdot S_{j} + J_{2}\sum_{i, j \in N(N(i))} S_{i} \cdot S_{j}
\end{align}

Note, $S_{i}$ and $S_{j}$ represent the total spin operator applied to site $i$ and $j$, respectively. The $N(i)$ represents the neighbors of site $i$ and $N(N(i))$ represents the next nearest neighbors of site $i$. The system being modeled is a spin system with nearest neighbor and next-nearest neighbor interactions weighted by the $J_{1}$ and $J_{2}$ magnetic coupling terms, respectively. For all systems considered, periodic boundary conditions were employed to help simulate the behaviors of a bulk material through approximating the infinite grid with a finite-sized approximation. The aim is to determine the phase transitions for these spin systems. 

The Heisenberg model is particularly valuable for studying various physical phenomena, such as the spin properties in layered materials \cite{feulner2022quantum}. It also enables the exploration of exotic phases like quantum spin liquids (QSLs), which can be studied with the \( J_{1}-J_{2} \) Heisenberg model. This model facilitates the investigation of long-range quantum entanglement in periodic quantum spin systems, which might be relevant for areas such as high-\( T_c \) cuprate superconductors, where QSL phases may be linked to superconductivity through low-energy meta-stable states \cite{anderson1987resonating, wen1989chiral, read1991large, lee2006doping, poilblanc2014resonating}. Additionally, QSLs may host topological excitations essential for topological quantum computing \cite{broholm2020quantum}.

Simulating the \( J_{1}-J_{2} \) model is challenging. Various conclusions have been drawn about the phases between the N\'eel phase and the collinear phase. These include the columnar valence-bond solid phase \cite{dagotto1989phase,sachdev1990bond}, the plaquette valence bond solid state \cite{mambrini2006plaquette}, and both the gapless and gapped quantum spin liquid (QSL) states \cite{jiang2012spin}. In the latest work by Liu, Gong, \textit{et. al.}, it was argued that for fixed \( J_{1} \), a N\'eel antiferromagnetic (AFM) phase exists for \( J_{2} \leq 0.45 \), a gapless QSL phase for \( 0.45 \leq J_{2} \leq 0.56 \), a valence bond solid (VBS) phase for \( 0.56 \leq J_{2} \leq 0.61 \), and a collinear AFM phase for \( J_{2} \geq 0.61 \) \cite{liu2020gapless}.

The current work expands on prior investigations of spin systems simulation using quantum computing. For instance, Lotshaw \textit{et. al.}'s investigation focused on the simulation of frustrated Ising Hamiltonians using the quantum approximate optimization algorithm (QAOA), specifically to  explore frustration in Shastry-Sutherland and triangular lattices \cite{lotshaw2023simulations}. Chowdhury \textit{et. al}. explored the isotropic Heisenberg model by leveraging trotterization for a time evolution study on a superconducting quantum computer. \cite{chowdhury2023enhancing} Our study chooses to explore the square $J_{1}-J_{2}$ Heisenberg model, with periodic boundary condition.

Chowdhury \textit{et. al}. described the practical aspect of simulation through the discussion of error mitigation strategies like Zero-Noise Extrapolation (ZNE) and dynamic decoupling, highlighting the importance of error mitigation in obtaining accurate simulations on NISQ devices. \cite{chowdhury2023enhancing} Uvarov \textit{et. al}, specifically explored VQE for frustrated quantum systems using the Hubbard model with nearest and next-nearest interactions using a quantum simulator of Rydberg atoms. \cite{uvarov2023variational} The work noted several key considerations for the successful implementation of VQE: (i) the meta-optimization of ansatz layers and application number, (ii) the optimization difficulty intrinsic to barren plateaus, and (iii) the development of accurate correlation function metrics for phase identification. 


Huerga lead a study similar to the current investigation using the square $J_{1}-J_{2}$ Heisenberg model. \cite{huerga2022variational} Huerga simulated the valence bond solid phase with a cluster-Gutzwiller ansatz based on Hierarchical Mean-Field Theory (HMFT). \cite{huerga2022variational} The study recovered the phase diagram via investigation of plaquette and magnetization correlation functions using the VQE method. \cite{huerga2022variational} Feulner and Hartmann, in a similar vein, proposed another ansatz for the system examining its size scaling and ability to reproduce $\text{X}$ correlation values. \cite{feulner2022quantum} Note, Feulner and Hartmann tackled a different problem for a system without periodic boundary conditions. Their optimization plateaued to an approximate ground state after $\sim 150,000$ VQE optimization steps. Inspiration for our ansatz and comparison with these prior theoretical studies is used to validate our study and to expand the discussion. \cite{huerga2022variational, feulner2022quantum} Our ansatz is a variant of TwoLocal, a parameterized circuit with alternating rotation and entanglement layers, in this case with CX entangling gates, RZ rotation gates, in a linear entanglement structure which we found to improve performance over prior ansatz by $9.31\%$. We differentiate ourselves from prior works by 1) switching our explorations from theoretical performance to experimental performance and 2) extending the simulation goal from capturing expected performance of specific phases within regions of the $J_{2}/J_{1}$ boundary to $0 \le J_{1}/J_{1} \le 1$.

Additionally, our experimental implementation of QLanczos for the spin system expands on prior theoretical work by Kirby, Motta, and Mezzacapo \cite{kirby2023quantum} by combining it with Variational Fast-Forwarding (VFF) \cite{filip2024variational}. The method allows us to implement the theoretical method experimentally on a 4-spin site instance. A computation study of the Quantum Computed Moments algorithm, a theoretical modification to the estimation produced by VQE based on Hamiltonian moments, is also performed for a 9-spin site instance, following the algorithm proposed by Vallury \textit{et. al}, in Appendix A. \cite{Vallury2020, Vallury2023}

\section{Methods}

\paragraph{} In this section, the specific choice of observables used to probe the $J_1-J_2$ system is described. The section describes the theory behind the algorithms employed, including warm-started VQE and the QLanczos approach. Finally, the section goes into the practical aspects of the experiments: zero-noise extrapolation, dynamical decoupling, zero-noise classical correction, block encoding representation of $\textbf{H}$, variational fast-forwarding, and run-time settings.

\subsection{Correlation Values}

\paragraph{} The presence of desirable phases can be detected by examining correlation expectation values. \cite{liu2020gapless} Key but insufficient properties indicating a quantum spin liquid phase include a high degeneracy, relatively strong global interaction, low classical magnetic ordering, and non-trivial global correlation-characteristics consistent with a QSL state. \cite{zhou2017quantum} For our purposes, a spike in the expectation value of the dimer correlation function, a greater global spin correlation $S_{corr, global}$ (see Eq.~~\ref{SG}) than local spin correlation, $S_{corr, local}$ (see Eq.~~\ref{SL}), and a low N\'eel order according to the analysis performed by Liu \textit{et. al}. \cite{liu2020gapless}.

The N\'eel AFM phase is described by the N\'eel AFM order, Eq.~~\ref{M0} \cite{liu2020gapless}: 
\begin{align}
  \label{M0}
   \left<M_{0}\right> = \frac{1}{N^{4}}\sum_{i,j} (-1)^{i_{x} + i_{y}}(-1)^{j_{x} + j_{y}} (\textbf{S}_{i} \cdot \textbf{S}_{j})
\end{align}

In this case, $M_{0}$ represents the expectation of total magnetization, $N$ denotes the number of sites, $S_{i}$ and $S_{j}$ represent the total spin operators, $i$ and $j$ stand for two sites of the lattice, $i_{x}$ denotes the $x$ index of $i$, and $i_{y}$ denotes the $y$ index of $i$.

\subsubsection{Dimer Order}

\paragraph{} The Valence Bonding state is characterized by its characteristic dimers, which are detected via the Horizontal Dimer correlation function, Eq.~ \ref{DX} \cite{liu2020gapless}. $D_{x}^{2}$ represents the horizontal dimer expectation squared and $e_{x}$ denotes a unit vector of $x$, taken to be the horizontal direction:

\begin{multline}
      \label{DX}
    \left<D_{x}^{2}\right> = \frac{1}{N(N - 1)}\sum_{i, j \in N(i)}(-1)^{i_{x} + j_{x}}\\(\textbf{S}_{i} \cdot \textbf{S}_{i + e_{x}}) \cdot (\textbf{S}_{j} \cdot \textbf{S}_{j + e_{x}})
\end{multline}

\subsubsection{Local/Global Z Correlation Values}

\paragraph{} In a similar choice to Liu \textit{et. al}.'s study, the spin correlation dynamic is probed using $Z$-component correlation functions \cite{liu1965correlation}:

\begin{align}
  \label{SL}
    \left<S_{corr, local}\right> = \frac{1}{4N}\sum_{i, j \in N(i)} (Z_{i} \cdot Z_{j})
\end{align}

\begin{align}
  \label{SG}
    \left<S_{corr, global}\right> = \frac{1}{N(N - 1)}\sum_{i, j} \frac{(Z_{i} \cdot Z_{j})}{|i_{x} - j_{x}| + |i_{y} - j_{y}|}
\end{align}

$S_{corr, local}$ represents the sum of \textbf{Z} Pauli correlation value between neighboring spin site $\textbf{Z}_{i}$ and $\textbf{Z}_{j}$ averaged relative to the number of sites considered. Likewise, $S_{corr, global}$ represents the averaged \textbf{Z} Pauli correlation value between a site and all other sites $\textbf{Z}_{i}$ and $\textbf{Z}_{j}$ averaged relative to the number of sites considered.

\subsubsection{Energy Evaluation}

\paragraph{} The energy of the resulting states is also explored by leveraging Eq.~\ref{ham}. Near $J_{2}/J_{1} = 0.5$, the domain with frustration, one would generally expect an increase in energy. The energy observable provides a means to assess the quality of optimization as an absolute and relative metric. As an exact metric, we compare our results with exact diagonalization, the ground truth.
We use the energy as a metric to study how the system's ground state energy varies as we tune the $J_{2}/J_{1}$ ratio, in addition to algorithmic details.
We can then inspect relative performance of each approach.


\subsection{Simulation Methods}

\subsubsection{Warm-Starting}

\paragraph{} For the problem of interest, with $Q$ spins, the classical intractability arises due to the infeasibility of simulating all $2^{Q}$ possible configurations. Given independent spins, the optimization problem would reduce from a $2^{Q} \times 2^{Q}$ linear operator minimization problem to a $2Q \times 2Q$ linear operator minimization problem. The independence allows the simplification by reducing the set of configurations to require only tracking the 2D state of $Q$ spins. In this case, the simulated annealing algorithm provided by Scipy is used to implement such an optimization routine and the averaged (4 different runs) parameter values is used to create initial qubit positions via a $U(\theta_{k}, \phi_{k})$ gate. This serves as a warm-start mechanism. More generally, the work extends a long line of warm-starting approaches to quantum computing tasks in the NISQ-era. \cite{harwood2022,truger2024warmstarting,beaulieu2021maxcut,truger2024,li_warm_2023} For more technical details, see the Appendix section A.

\subsubsection{VQE}

\paragraph{} VQE is an algorithm for ground state simulation on near-term devices consisting of four main components: conversion of the problem Hamiltonian to a spin Hamiltonian, parameterized ansatz state preparation, Hamiltonian expectation calculation, and classical optimization of the parameterized circuit's Hamiltonian expectation value. To create a bijection, one assigns the sites described by the problem directly to qubits on the quantum computer. In this case, the spins can be trivially bijected since each site can be described by a spin-$1/2$ state. The quantum computer's state is then parameterized implicitly through the choice of ansatz by its associated parameters. By optimizing the analogous spin state to a state with similar energy expectation to the system of interest, one produces a state that simulates the ground state for the system of interest. 

\subsubsection{QLanczos}

\paragraph{} QLanczos is a simulation method based on the Lanczos method in classical computing and is, in general, a type of quantum subspace diagonalization method \cite{kirby2023quantum}. Instead of diagonalizing the Hamiltonian directly, one queries both its action on a subspace, called the support space, to generate a Hamiltonian $\textbf{H}$ and inner product of the basis vectors to generate a similarity matrix, $\textbf{S}$. One then solves the following generalized eigenvalue problem:

\begin{align}
  \label{krylov}
    \textbf{H}\ket{c} = \lambda \textbf{S}\ket{c}
\end{align}

The generated $\ket{c}$ describes the coefficients determining a linear combination in the support space which, ideally, approximates the desired solution. To ensure convergence, one requires an initial state with $\Omega(1/\text{poly}(n))$ overlap with the ground state. \cite{yeter2021,Tkachenko_2024,10.1007/978-981-15-9927-9_27,PRXQuantum.2.010333,Nandy:2024htc} The choice of basis can vary from real-time evolution to the Lanczos basis. However, the Lanczos basis is particularly promising for the system of interest according to prior literature. \cite{kirby2023quantum} The insight provided by the prior work is the creation of an efficient and exact method that uses quantum signal processing (QSP) to encode the Chebyshev polynomials following the standard LCU method. \cite{kirby2023quantum} The encoding spans the same subspace as the Lanczos basis provides, and has desirable characteristics like a $O(n)$ query cost to determine $\textbf{H}$ and $\textbf{S}$ unlike the naive implementation of $O(n^{2})$. \cite{kirby2023quantum}

The method introduced by Kirby, Motta, and Mezzacapo defines the $\textbf{H}$ and $\textbf{S}$ matrix in terms of $\textbf{T}_{k}(H)\ket{\psi_{0}}$ for $\ket{\psi_{0}}$ where $\textbf{T}_{k}$ represents the $k$th  Chebyshev polynomials. To see details on how one computes the matrix elements from expectation values derived from quantum circuits, see Appendix section C. Note, we deivate from Kirby, Motta, and Mezzacapo by defining a block encoding construction of $\textbf{U}$ is handled through through the use of FABLE, a fast approximate quantum circuit construction library for block encoding. \cite{kirby2023quantum, camps2022fable} Naively, a perfect fidelity reconstruction of the originally proposed technique results in a circuit depth of $\sim 1500$ for the Rensselear system. But, with a modest fidelity reduction of $0.01\%$, we can reduce the depth greatly to $\sim 1000$.

\subsection{Experimental Settings}

\subsubsection{Experimental Device} 

\paragraph{} The quantum computer used to experimentally perform the optimization was the IBM Rensselear, a 127-qubit IBM Eagle V3 processor. Exact diagonalization was performed with NumPy's diagonalization functionality to compute the theoretical "Classical" results via Qiskit's Scipy eigensolver.

\subsubsection{Zero Noise Extrapolation (ZNE)}

\paragraph{} To implement algorithms like VQE, one requires accurate expectation values to be produced by the quantum computer. Zero noise extrapolation (ZNE) is a technique used to mitigate gate errors on near-term quantum computers to achieve better quality estimates of expectation values. It works by changing a quantum program to run at different noise levels, by extending circuit depth through methods like circuit unrolling, to generate different estimates for expectation values of a given observable. It then estimates the true observable by extrapolating the changes to the observable with the given finite, non-zero noise levels to that of the noiseless case.

\subsubsection{Dynamic Decoupling}
 
\paragraph{} Dynamical decoupling (DD) is a hardware technique in quantum computing that uses time-dependent control modulation to suppress decoherence and protect qubits from the environment. \cite{Yang2011, PhysRevLett.108.086802, Szwer_2011, PhysRevApplied.16.054047} DD uses a periodic sequences of control pulses that average the unwanted system-environment coupling to zero. \cite{Yang2011, PhysRevLett.108.086802, Szwer_2011, PhysRevApplied.16.054047}

\subsubsection{Classically-Reinforced ZNE}

\paragraph{} ZNE assumes noise is uncorrelated with the Hamiltonian and, thus, its impact is mainly a function of time. However, to do this, one requires a function fit to capture the observable change relative to exposure to the error. But, the extrapolated function itself may have an incorrect zero-intercept, $d$, an unknown scaling between noise-level and true time, $c$, and a constant, systemic error in estimating first derivatives, $b$, and second derivatives, $a$. Thus, it is assumed instead one is observing a quadratic transformation of the original data, Eq.~\ref{adapted_zne}. With this assumption, one may ask if such a transformation is invertible in hopes of generating the true correction. Desirable properties of inverting the transformation are (i) the gradients match as closely as possible and (ii) the function closely matches given references values. Once these parameters are found, $f_{\text{ZNE}}$, our ZNE corrected inputs, can be given and $f_{\text{True}}$, a more reliable estimate for the observable, will be generated. The Scipy optimization library with Nedler-Mead is used with the objective of finding a transformation with the above properties.

\begin{align}
  \label{adapted_zne}
    f_{\text{True}} = a(\frac{f_{\text{ZNE}}(x)}{c})^{2} + b(\frac{f_{\text{ZNE}}(x)}{c}) + d
\end{align}

The primary objective is to minimize the squared difference of the first derivatives and second derivatives of the transformed observable with the numerical first and second derivative estimates of the original data. Violation to ensure point matching is allowed by a term in the objective function being weighted by a penalty factor $\beta = \frac{1}{100}$. However, to find the inversion, as described previously, reference values must exist.

\subsubsection{Variational Fast-Forwarding}

\paragraph{} A significant issue with the naive implementation of the necessary operators for QLanczos is the necessity to perform the block encoding containing $\textbf{H}$. Another issue is the efficient implementation of the reflection operator, requiring reflection along the weighted superposition of the involved Pauli terms composing $\textbf{H}$, weighted by the coefficient associated with the term. This process has to be done $k$ times in order to extract the $k$th basis vector, since one needs to produce $\textbf{H}^{k}$. This is particularly an issue as the block encoding itself takes about $\sim 1000$ gate operations. An approximate operation is desirable. \cite{PhysRevA.106.042409, filip2024variational} A great reduction in depth can occur if one can find an $\textbf{A}$ such that $\textbf{A} = \textbf{U}\textbf{D}(\overrightarrow{\gamma})\textbf{U}^{\dagger}$. In this case, one is implicitly finding, by a variational approach, a SVD decomposition. Note, $\overrightarrow{\gamma}$ representing the vector of the diagonal weights. The utility comes about when applying the operator $\textbf{A}$ $k$ times where the SVD nature can be used to derive the following: $\textbf{A}^{k} = \textbf{U}\textbf{D}(k\overrightarrow{\gamma})\textbf{U}^{\dagger}$. Thus, the decomposition allows a fixed depth circuit construction to represent the operator $\textbf{A}^{k}$ for any positive, non-zero application number $k$. For our study, we follow the implementation of Filip, Ramo, and Fitzpatrick \cite{filip2024variational}. See Appendix section B for technical details on the method's implementation.

\subsubsection{Settings}

\paragraph{} To run the experiments, zero-noise extrapolation with factors [1, 1.25, 1.5, 1.75, 2.0] was used, with 5000 shots, XX dynamic decoupling, level 2 transpilation, and readout circuit twisting. The results are averaged expectation over 3 different sets of correlation measurements. 

\section{Results}

\paragraph{} The results demonstrate that, despite the inherent noise present in current NISQ-era hardware, it is possible to capture the significant trends of the $J_{1}-J_{2}$ Heisenberg model using VQE, particularly in the identification of phase transitions, if one employs modern mitigation techniques, a novel computation not done in prior works. We see that complementing VQE with classical correlation values when available to correct estimates yields significantly better quantitative agreement.

The theoretical performance of QLanczos proves to be promising for extracting observables while its experimental implementation proves to face limitations mainly through the efficient near-term construction of the necessary operators for the QLanczos method. Although, improvements to the efficiency of variational fast-forwarding or hardware improvements are likely to significantly lessen these obstacles.

\subsection{Warm-Starting}

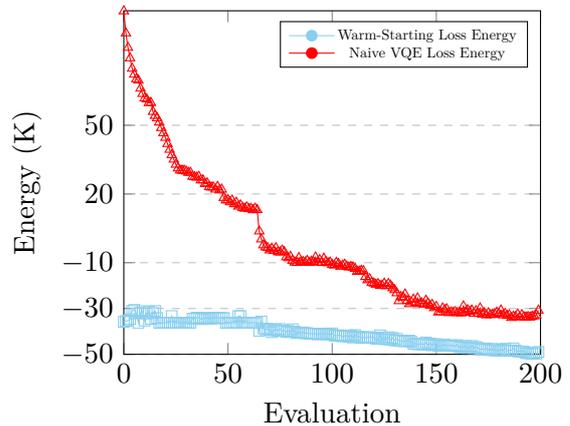
\begin{figure}
\centering
\noindent\begin{tikzpicture}
\begin{axis}[
    xlabel={Evaluation},
    ylabel={Energy (K)},
    xmin=0, xmax=200,
    ymin=-50, ymax=100,
    xtick={0,50, 100,150,200},
    ytick={-50, -30, -10, 20, 50},
    legend pos=north east,
    ymajorgrids=true,
    grid style=dashed,
    scale=0.8, 
    legend style={nodes={scale=0.5, transform shape}}, 
    legend image post style={mark=*}
]

\addplot[
    color=babyblue,
    mark=square,
    ]
    coordinates {
    (0,-35.8399999981978)(1,-35.49841065207264)(2,-35.2400497689767)(3,-34.63939942567898)
    (4,-31.0615118135097)(5,-30.58558138365101)(6,-32.137105344950434)(7,-36.237295970382206)
    (8,-31.924383731000226)(9,-31.141628002954555)(10,-35.302990634173405)(11,-32.223973319823216)
    (12,-31.398850930537833)(13,-34.52282666392236)(14,-31.91384906765184)(15,-31.050042441730085)
    (16,-34.17784210136995)(17,-36.23729597038219)(18,-36.23729597038219)(19,-36.23729597038219)
    (20,-36.23729597038219)(21,-36.23729597038219)(22,-36.23729597038219)(23,-33.290083138212715)
    (24,-36.23729597038219)(25,-36.23729597038219)(26,-36.23729597038219)(27,-36.23729597038219)
    (28,-36.23729597038219)(29,-36.23729597038219)(30,-36.23729597038219)(31,-36.23729597038219)
    (32,-36.23729597038219)(33,-34.41450898731283)(34,-34.41452213803742)(35,-33.9919032494419)
    (36,-33.99187448904562)(37,-34.60052448461948)(38,-34.60056029529795)(39,-34.08654569778305)
    (40,-34.086537282285335)(41,-34.414517534777886)(42,-34.41448775431072)(43,-33.9919266039872)
    (44,-33.9918824671404)(45,-34.17785030067148)(46,-34.17783526548563)(47,-34.17781926120636)
    (48,-34.17784210136995)(49,-36.23729597038219)(50,-36.23729597038219)(51,-36.23729597038219)
    (52,-36.23729597038219)(53,-36.23729597038219)(54,-36.23729597038219)(55,-33.290083138212715)
    (56,-33.290083138212715)(57,-36.23729597038219)(58,-36.23729597038219)(59,-36.23729597038219)
    (60,-36.23729597038219)(61,-36.23729597038219)(62,-36.23729597038219)(63,-36.23729597038219)
    (64,-36.23729597038219)(65,-39.746534247688494)(66,-35.649460972211735)(67,-38.74032301567574)
    (68,-39.14798732429631)(69,-38.84666184933564)(70,-39.29815870195063)(71,-40.344117994788625)
    (72,-39.590577828944056)(73,-39.05380745403698)(74,-39.66872199089934)(75,-38.96723502696749)
    (76,-39.85494268493854)(77,-39.602843071387866)(78,-40.279566726702676)(79,-39.79186798905436)
    (80,-40.27265835057549)(81,-39.07698668585604)(82,-40.07953909834606)(83,-40.75429068231495)
    (84,-40.605876951004895)(85,-40.62610746253956)(86,-40.41298948584349)(87,-40.796167821512846)
    (88,-40.73299974067896)(89,-41.06143635962609)(90,-40.97125743002205)(91,-41.044485967015746)
    (92,-40.946009405451655)(93,-41.328256447389435)(94,-41.16106180283245)(95,-40.927630029132956)
    (96,-41.05435800126536)(97,-41.186808610781085)(98,-41.2108129335262)(99,-41.40616893722948)
    (100,-40.855164700172885)(101,-41.82063029548913)(102,-41.94567610879585)(103,-41.70175409927402)
    (104,-41.742866208231945)(105,-42.24545935658921)(106,-41.91845208889549)(107,-41.89368827280098)
    (108,-42.285494056723564)(109,-41.77600663168359)(110,-42.69866539255141)(111,-41.5735772058369)
    (112,-42.28065869748483)(113,-42.35744647258253)(114,-42.88959518510184)(115,-42.51539225930438)
    (116,-42.809288438712834)(117,-42.6031109001339)(118,-42.57389857033335)(119,-42.59833768865002)
    (120,-42.72413517350414)(121,-43.17160956175368)(122,-43.29694090100793)(123,-41.70429829419494)
    (124,-43.33347191852403)(125,-42.622937364234986)(126,-43.45513603459094)(127,-42.56155578414092)
    (128,-43.61957475889253)(129,-42.8004649057596)(130,-43.95446078845432)(131,-43.261205747969655)
    (132,-44.09830679057507)(133,-44.17256900136081)(134,-44.293423047910544)(135,-44.227774380190205)
    (136,-44.093768525313)(137,-44.00858915526063)(138,-44.50451014098417)(139,-45.39044706541162)
    (140,-45.46355772411531)(141,-45.64332682507812)(142,-45.503049408067675)(143,-45.53919214925507)
    (144,-45.7433512053367)(145,-44.82702109718253)(146,-45.89309944574488)(147,-45.288470988028145)
    (148,-46.04570451765772)(149,-45.559543907752925)(150,-45.63364852180897)(151,-46.303298046985816)
    (152,-46.2839613982113)(153,-46.18639132874191)(154,-45.66625318313203)(155,-46.24072923851493)
    (156,-46.30093901232481)(157,-46.06644914144851)(158,-46.24235794886879)(159,-46.45393014665038)
    (160,-45.96776276383269)(161,-46.96331920440325)(162,-46.779969921510016)(163,-46.95188917491059)
    (164,-46.281063363489054)(165,-47.28085276252512)(166,-47.17143630831241)(167,-46.33801222965599)
    (168,-47.30096307873126)(169,-46.8141259648567)(170,-47.22930106573936)(171,-47.850459079757194)
    (172,-47.714324532175965)(173,-47.95928488535977)(174,-47.40809575195608)(175,-48.11649477512111)
    (176,-48.15911553860419)(177,-47.59327868003679)(178,-48.29287882593168)(179,-48.32390905340935)
    (180,-48.459489354660924)(181,-47.90908774594769)(182,-48.33738601403015)(183,-47.53732398807964)
    (184,-48.515294803965645)(185,-48.340199702738296)(186,-48.2436756035998)(187,-47.39174478348079)
    (188,-48.3273695248023)(189,-48.73524330985279)(190,-48.77890435296118)(191,-48.77869877411865)
    (192,-48.800314024649055)(193,-49.481219447715695)(194,-49.481096208864166)(195,-49.50200014270767)
    (196,-49.85508503804886)(197,-49.38192548837247)(198,-49.74145672955279)(199,-48.9446505368389)
    };
    \addlegendentry{Warm-Starting Loss Energy} 

    \addplot[
    color=red,
    mark=triangle,
    ]
    coordinates {
    (0,99.84000000000003)(1,90.20473633099621)(2,83.9087720530891)(3,79.29409951592619)
    (4,74.78724543409385)(5,72.19450023660745)(6,70.16481504395178)(7,69.58624190391585)
    (8,66.10771232927681)(9,63.76622720986248)(10,61.80563537759087)(11,61.354941314941215)
    (12,59.75397347967103)(13,59.82560616545923)(14,55.933949884714316)(15,54.18435081824563)
    (16,53.15324858941497)(17,51.301967188321775)(18,48.8840262248105)(19,46.58717096781418)
    (20,44.59859285361834)(21,41.91278292586226)(22,39.384624288244595)(23,37.0131789827764)
    (24,35.11005372595208)(25,32.8561382434377)(26,31.24658357514158)(27,30.35368769431657)
    (28,30.242958143845772)(29,30.24295814384578)(30,29.887315531447467)(31,29.339518395505383)
    (32,29.33951839550538)(33,27.613949676444708)(34,27.411380879836013)(35,27.19666162121624)
    (36,27.596569064490858)(37,26.020228530862923)(38,26.05755983279862)(39,24.462906949913936)
    (40,24.73929864729361)(41,23.094300957730873)(42,22.907362352706873)(43,22.92791319882329)
    (44,22.013653680966147)(45,23.193390823621616)(46,21.620402974913116)(47,22.001488876023842)
    (48,18.380024450249167)(49,17.25467886805366)(50,17.37146706286449)(51,16.47746080550665)
    (52,17.225525186267618)(53,15.629756951677452)(54,16.269397754104396)(55,14.5918102994602)
    (56,15.12485543525544)(57,13.959783470788617)(58,13.664629017028073)(59,14.233706538036898)
    (60,13.465390968279248)(61,13.626245864775154)(62,12.985766027898832)(63,13.673880481289318)
    (64,13.071176751176585)(65,3.7438708621689347)(66,0.245758782657698)(67,-2.320983377312833)
    (68,-3.226261069939569)(69,-3.5013334478732543)(70,-4.547195556078666)(71,-3.8549253841924918)
    (72,-4.73338296607568)(73,-3.5935864934911037)(74,-5.357841168730027)(75,-5.470575584312679)
    (76,-4.619560363198688)(77,-5.259374108934299)(78,-7.536740063639142)(79,-7.955834096652596)
    (80,-7.379587198543244)(81,-9.053006464926824)(82,-9.85939474913691)(83,-10.011680342369218)
    (84,-7.8792970967361144)(85,-9.823275800043161)(86,-9.641165753945115)(87,-9.879408195792864)
    (88,-9.143325678885498)(89,-9.97093684746979)(90,-9.46928850045825)(91,-9.699501361553885)
    (92,-7.63357598230673)(93,-9.85096028859795)(94,-8.779766708712366)(95,-10.037282469094418)
    (96,-8.211883742477907)(97,-9.584155712266984)(98,-10.162482003138031)(99,-9.906345471234367)
    (100,-10.887504477578652)(101,-11.175354198433705)(102,-9.971753676092495)(103,-10.445079716667419)
    (104,-11.674779287949956)(105,-11.206437530574865)(106,-10.816512508380372)(107,-11.675755259224397)
    (108,-12.007145608533776)(109,-11.793879704005311)(110,-12.033190713997733)(111,-12.629814894276045)
    (112,-13.897789173850546)(113,-13.824232088011751)(114,-13.466869499465055)(115,-13.611093696101243)
    (116,-16.153774811800258)(117,-17.577402603374804)(118,-17.132468626112185)(119,-18.67060575578387)
    (120,-19.005213271732945)(121,-19.52824517475462)(122,-19.292324378386912)(123,-20.06197946846779)
    (124,-19.972893160798325)(125,-19.091077353382698)(126,-20.20498081941637)(127,-19.219282650966232)
    (128,-19.886184243640464)(129,-21.089349829163883)(130,-21.899283306394572)(131,-23.260886929219605)
    (132,-26.725025563492594)(133,-24.906534492854064)(134,-25.170333165508232)(135,-23.780835767973848)
    (136,-26.598904106175823)(137,-28.03170700731741)(138,-27.192746414726045)(139,-26.066285669233995)
    (140,-27.53992415723586)(141,-25.487727657756572)(142,-28.048878992367552)(143,-27.652898095644435)
    (144,-28.21007116171184)(145,-26.53342373174953)(146,-27.75404171963356)(147,-28.80842450959354)
    (148,-28.344044188397856)(149,-29.771910425911887)(150,-29.20839145184558)(151,-31.642827353982742)
    (152,-31.512307586967957)(153,-29.87534879484528)(154,-30.640761742968238)(155,-30.275933705713868)
    (156,-31.92288767986186)(157,-31.619779562094113)(158,-31.7748195050287)(159,-31.499474788711844)
    (160,-31.95047485927418)(161,-31.860419675340573)(162,-32.044296390764885)(163,-29.0189011715676)
    (164,-31.782657881701873)(165,-30.65836498265397)(166,-31.38701150011781)(167,-30.84265612288331)
    (168,-32.15095772387185)(169,-31.085486714943336)(170,-32.156966403122325)(171,-29.884914537681578)
    (172,-32.596488668054825)(173,-31.93325193150622)(174,-32.91408257865997)(175,-31.987281480192664)
    (176,-32.78972426826348)(177,-32.68763200179563)(178,-32.501866296938076)(179,-31.77196465875421)
    (180,-32.290596989704724)(181,-30.803534444170985)(182,-32.93646011604036)(183,-32.29274278832257)
    (184,-33.116967333451505)(185,-32.87274084473037)(186,-34.05229983712483)(187,-33.48345521986816)
    (188,-33.678785883202956)(189,-32.5191685781237)(190,-33.81799810577603)(191,-32.817683087418416)
    (192,-33.65614262846305)(193,-33.850023608260635)(194,-33.671927598800394)(195,-33.86341717775096)
    (196,-33.48592535134583)(197,-33.34893255687793)(198,-32.901333727460205)(199,-30.821037023819514)
    };
    \addlegendentry{Naive VQE Loss Energy} 

\end{axis}
\end{tikzpicture}
\caption{Naive vs Warm-Starting Energy expectation versus VQE iteration for $J_{2}/J_{1} = 0.56$, performed with the EfficientSU2 ansatz (1 repetition) and using NFT optimizer.}
\label{warm_opt}
\end{figure}

\paragraph{} We consider applying warm-starting, a method typically reserved to quantum computation of classical optimization problems, to the problem of generating a good initial state to initialize the VQE optimization. \cite{harwood2022,truger2024warmstarting,beaulieu2021maxcut,truger2024,li_warm_2023} To explore the effects of warm-starting, two trial runs of VQE were performed, with one using warm-starting and one without it. Both trials used a one layer EfficientSU2 ansatz at the $J_{2}/J_{1} = 0.56$ point, the point of expected maximal frustration on a noiseless simulator. The variance of the warm-starting approach is $19.87$ while the standard deviation is $4.45$ compared with zero variance and standard deviation for the random initialization. As one can see in Figure \ref{warm_opt}, warm-starting optimization poses a significant benefit over random parameter settings. The results are intriguing because it shows that warm-starting can prove beneficial even when the assumption being used for the initialization is far from accurate. When examining the parameter results of warm-starting, it becomes clear that the AFM nature is being captured as the parameters tend to either be near $0$, $\pi$, and $2\pi$. Such an assumption is confirmed when trialing an Ising model approximation and seeing similar performance gains. The warm-starting technique was employed for all data collected.

\begin{figure}
\begin{tikzpicture}
\begin{axis}[
    ybar,
    bar width=1.0pt, 
    scale=0.8,
    ymin=-70, ymax=-30,
    ylabel={Energy (K)},
    enlarge x limits=0.08, 
    xtick=data,
    xticklabels={},
    x tick label style={rotate=45, anchor=east, align=right, text width=0.6cm}, 
    legend style={
        at={(0.3,-0.30)}, 
        anchor=north,
        legend columns=-1,
        /tikz/every even column/.append style={column sep=0.1cm}, 
        text width=1.2cm, 
        draw=none,
        font=\tiny 
    },
    ylabel near ticks,
    ymajorgrids=true,
    grid style=dashed,
    nodes near coords,
    every node near coord/.append style={font=\tiny, anchor=north}, 
]

\addplot coordinates {(1, -55) (2, -52) (3, -53)};
\addlegendentry{TwoLocal}

\addplot coordinates {(4, -47) (5, -33) (6, -36)};
\addlegendentry{Real}

\addplot coordinates {(7, -38) (8, -34) (9, -32)};
\addlegendentry{Eff. SU2}

\addplot coordinates {(10, -48) (11, -46) (9, -47)};
\addlegendentry{HMFA}

\addplot coordinates {(11, -34) (12, -36) (13, -37)};
\addlegendentry{Feulner}

\draw [red, dashed] ({rel axis cs:0,0}|-{axis cs:1,-66}) -- ({rel axis cs:1,0}|-{axis cs:1,-66}) node [midway, above] {Ground Truth};

\end{axis}
\end{tikzpicture}
\caption{Comparison of VQE Simulation Methods (2-Local, Real, EfficientSU2, HMFA, and Feulner) over 3 runs at $J_{2}/J_{1} = 0.56$ with number of NFT iterations capped at 500 on a noiseless simulator.}
\label{vqe_ansatz}
\end{figure}
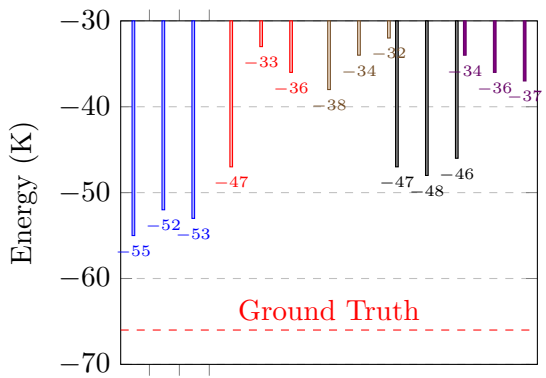

\subsection{VQE Method}

\paragraph{} We show the comparison of the following ansatz: TwoLocal, RealAmplitudes, HFMA, and Feulner in Figure \ref{vqe_ansatz}. One can see that the TwoLocal ansatz, defined with RX and CX gates, maintains a low variance when compared with RealAmplitudes and EfficientSU2 while yielding a better energy value than any other ansatz choice tested. Why does TwoLocal produces a higher quality result when compared to other more standard choices like RealAmplitudes or EfficientSU2? The prior study of the system performed by Huerga provides some insight wherein Hierarchical Mean-Field Theory (HMFA) predicted an  ansatz containing ZZ, XY, and Z rotations with a N\'eel starting state. \cite{huerga2022variational} Thus, it may not be surprising that the TwoLocal RX and CZ ansatz performed well, as it mirrors the HMFA model. 

The difference in convergence to an approximate ground truth energy between a designed ansatz like HFMA and a generic ansatz like RealAmplitudes or EfficientSU2 illustrates the potential benefits of using a theory-inspired ansatz for spin system simulation. 
\begin{figure}[htbp]
\begin{minipage}{0.49\textwidth}
\begin{tikzpicture}
\begin{axis}[
    xlabel={$J_{2}/J_{1}$ },
    ylabel={Correlation Function},
    xmin=0, xmax=1,
    ymin=-1.5, ymax=1.0,
    legend pos=north east,
    ymajorgrids=true,
    grid style=dashed,
    width=\textwidth,
    height=6cm,
    legend style={nodes={scale=0.5, transform shape}}
]
\addplot[color=babyblue, mark=square,] coordinates {
    (0,-0.7017802005268026)(0.1,-0.700895102070927)(0.2,-0.6971621329415036)(0.3,-0.6874489287103996)(0.5,-0.6053182704365818)(0.56,-0.5283801649185245)(0.58,-0.4884064509773419)(0.7,-0.16847269275029914)(0.8,-0.08564425664057519)(0.9,-0.059269946065780885)(1.0,-0.04639692549529759)
};
\addlegendentry{Classical Local Z}
\addplot[color=red, mark=square,] coordinates {
    (0,-0.6048370236562555)(0.1,-0.607037685841411)(0.2,-0.6079479627385287)(0.3,-0.6056843140065034)(0.5,-0.5593346760799276)(0.56,-0.5042545724970893)(0.58,-0.47371994933616957)(0.7,-0.2154575517425132)(0.8,-0.14657844173225357)(0.9,-0.12421851626803389)(1.0,-0.11315498398359748)
};
\addlegendentry{Classical Global Z}

\draw [babyblue] (axis cs:0.45,-1.5) -- (axis cs:0.45,1);
\addlegendimage{line legend, color=babyblue}
\addlegendentry{Neel/Gapless Boundary}

\draw [red] (axis cs:0.56,-1.5) -- (axis cs:0.56,1);
\addlegendimage{line legend, color=red}
\addlegendentry{Gapless/VBS}

\draw [black] (axis cs:0.61,-1.5) -- (axis cs:0.61,1);
\addlegendimage{line legend, color=black}
\addlegendentry{VBS/collinear}

\end{axis}
\end{tikzpicture}
\end{minipage}
\hfill
\begin{minipage}{0.49\textwidth}
\begin{tikzpicture}
\begin{axis}[
    xlabel={$J_{2}/J_{1}$ },
    ylabel={Correlation Function},
    xmin=0, xmax=1,
    ymin=-1.5, ymax=1.0,
    legend pos=north west,
    ymajorgrids=true,
    grid style=dashed,
    width=\textwidth,
    height=6cm,
    legend style={nodes={scale=0.5, transform shape}}, 
]
\addplot+[
    color=babyblue,
    mark=triangle,
    error bars/.cd,
    y dir=both, y explicit,
    ] coordinates {
    (0, -1.35)
    (0.2, -1.302) +- (0, 0.07123499494265068)
    (0.4, -1.05) +- (0, 0.07474729759897786)
    (0.5, -0.89) +- (0, 0.07745554902471198)
    (0.58, -0.36) +- (0, 0.08726594236681558)
    (0.8, -0.20) +- (0, 0.08800509052656447)
    (1.0, -0.05)
};
\addlegendentry{Experimental VQE Local}

\addplot+[
    color=red,
    mark=triangle,
    error bars/.cd,
    y dir=both, y explicit,
    ] coordinates {
    (0, -1.05)
    (0.2, -1) +- (0, 0.07973302438230963)
    (0.4, -0.88) +- (0, 0.08422647053078887)
    (0.5, -0.80) +- (0, 0.08730314090673491)
    (0.58, -0.36) +- (0, 0.09646086215639987)
    (0.8, -0.235) +- (0, 0.09845924806517817)
    (1.0, -0.1)
};
\addlegendentry{Experimental VQE Global}

\addplot+[green, mark=square, smooth]
    coordinates {
    (0.0, -0.9419)
    (0.1, -0.8556)
    (0.2, -0.8909)
    (0.3, -0.6009)
    (0.5, -0.4605)
    (0.56, -0.4042)
    (0.58, -0.34246)
    (0.7, -0.2423)
    (0.8, -0.1630)
    (0.9, -0.1536)
    (1.0, -0.1076)
};
\addlegendentry{Theoretical VQE Local Correlation}

\addplot+[
    color=orange,
    mark=square,
    smooth,
    error bars/.cd
    ]
    coordinates {
    (0.0, -0.7329)
    (0.1, -0.674)
    (0.2, -0.707)
    (0.3, -0.509)
    (0.5, -0.4319)
    (0.56, -0.3805)
    (0.58, -0.3266)
    (0.7, -0.243)
    (0.8, -0.18025)
    (0.9, -0.16449)
    (1.0, -0.13186)
    };
    \addlegendentry{Theoretical VQE Global Correlation}

\draw [babyblue] (axis cs:0.45,-1.5) -- (axis cs:0.45,1);
\addlegendimage{line legend, color=babyblue}
\addlegendentry{Neel/Gapless Boundary}

\draw [red] (axis cs:0.56,-1.5) -- (axis cs:0.56,1);
\addlegendimage{line legend, color=red}
\addlegendentry{Gapless/VBS}

\draw [black] (axis cs:0.61,-1.5) -- (axis cs:0.61,1);
\addlegendimage{line legend, color=black}
\addlegendentry{VBS/collinear}

\end{axis}
\end{tikzpicture}
\end{minipage}
\caption{Comparison of Exact Diagonalization with VQE Experiment across different $J_{2}/J_{1}$ choices. Top Panel: Theoretical results from diagonalization. Bottom Panel: Experimental and noiseless VQE simulation. Dashed horizontal lines indicate $0$ and $-1$, zero correlation and full anti-correlation, respectively.  Dashed vertical lines denote N\'eel/Gapless (babyblue), Gapless/VBS (red), and VBS/collinear (black) phase boundaries.}
\label{vqe_correlation}
\end{figure}
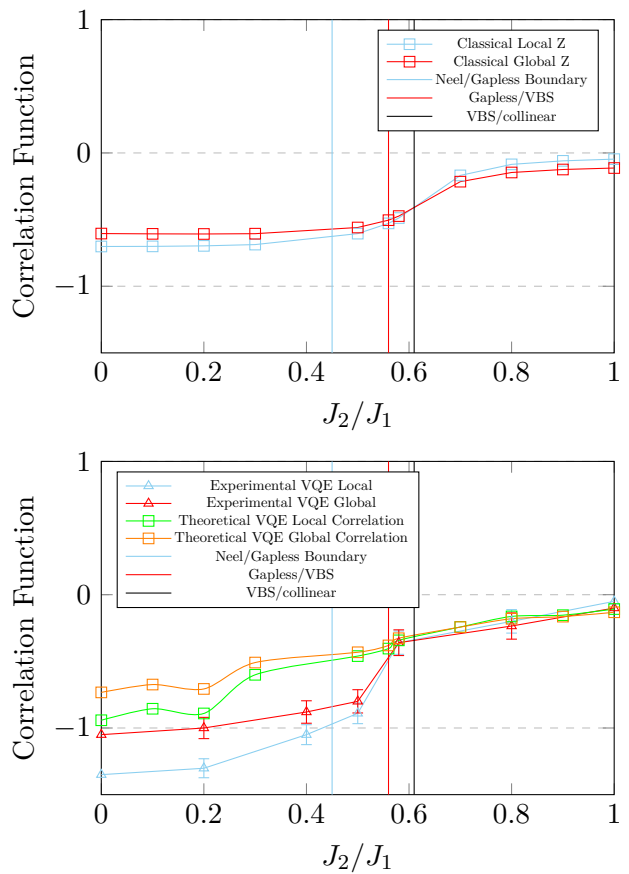

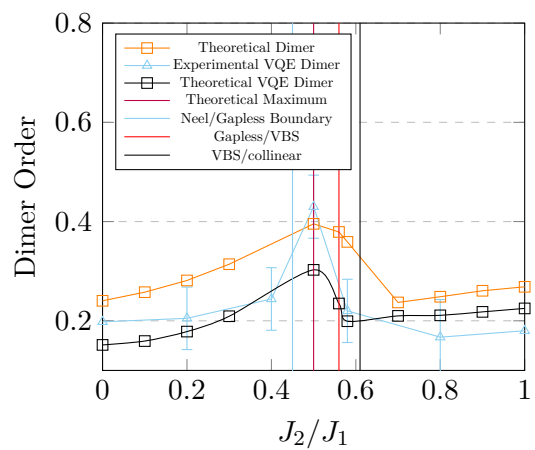
\begin{figure}[htbp]
\centering
\begin{tikzpicture}
\begin{axis}[
    xlabel={$J_{2}/J_{1}$ },
    ylabel={Dimer Order},
    xmin=0, xmax=1,
    ymin=0.1, ymax=0.8,
    legend pos=north west,
    ymajorgrids=true,
    grid style=dashed,
    scale=0.85,
    height=7cm,
    legend style={nodes={scale=0.5, transform shape}}, 
]

\addplot[
    color=orange,
    mark=square,
    ]
    coordinates {
    (0,0.24034282315311367)(0.1,0.2577057812956596)(0.2,0.2812798009644143)(0.3,0.3140678064223895)(0.5,0.39541050358144636)(0.56,0.3790479338032299)(0.58,0.3590460781380115)(0.7,0.23702407576179274)(0.8,0.24848603571885516)(0.9,0.26052818807036077)(1.0,0.26852262914756325)
    };
    \addlegendentry{Theoretical Dimer}

\addplot+[
    color=babyblue,
    mark=triangle,
    error bars/.cd,
    y dir=both, y explicit,
    ]
    coordinates {
    (0, 0.198)
    (0.2, 0.205) +- (0, 0.06297267715405934)
    (0.4, 0.244) +- (0, 0.0630647111065697)
    (0.5, 0.43) +- (0, 0.06356087185413259)
    (0.58, 0.22) +- (0, 0.06371246691767331)
    (0.8, 0.167) +- (0, 0.07585223660632465)
    (1.0, 0.18)
    };
    \addlegendentry{Experimental VQE Dimer}

\addplot[black, mark=square, smooth]
    coordinates {
    (0.0, 0.1514)
    (0.1, 0.1592)
    (0.2, 0.1783)
    (0.3, 0.2092)
    (0.5, 0.30267)
    (0.56, 0.2349)
    (0.58, 0.1992)
    (0.7, 0.2099)
    (0.8, 0.211)
    (0.9, 0.218)
    (1.0, 0.225)
    };
    \addlegendentry{Theoretical VQE Dimer}

\draw [purple] (axis cs:0.5,0.0) -- (axis cs:0.5,0.8);
\addlegendimage{line legend, color=purple}
\addlegendentry{Theoretical Maximum}

\draw [babyblue] (axis cs:0.45,0.0) -- (axis cs:0.45,0.8);
\addlegendimage{line legend, color=babyblue}
\addlegendentry{Neel/Gapless Boundary}

\draw [red] (axis cs:0.56,0.0) -- (axis cs:0.56,0.8);
\addlegendimage{line legend, color=red}
\addlegendentry{Gapless/VBS}

\draw [black] (axis cs:0.61,0.0) -- (axis cs:0.61,0.8);
\addlegendimage{line legend, color=black}
\addlegendentry{VBS/collinear}

\end{axis}
\end{tikzpicture}
\caption{Comparison of Theoretical and Experimental Dimer Orders across different $J_{2}/J_{1}$ choices. Dashed vertical lines denote N\'eel/Gapless (babyblue), Gapless/VBS (red), and VBS/collinear (black) phase boundaries.}
\label{vqe_dimer}
\end{figure}

\begin{figure}
\centering
\begin{tikzpicture}
\begin{axis}[
    xlabel={$J_{2}/J_{1}$ },
    ylabel={N\'eel Order},
    xmin=0, xmax=1,
    ymin=0, ymax=1.5, 
    legend pos=north east,
    ymajorgrids=true,
    grid style=dashed,
    scale=0.8,
    legend style={nodes={scale=0.5, transform shape}} 
]

\addplot+[purple, mark=triangle, error bars/.cd, y dir=both, y explicit]
    coordinates {
    (0.0, 0.60159) +- (0, 0.15)
    (0.1, 0.53101) +- (0, 0.15)
    (0.2, 0.54659) +- (0, 0.1)
    (0.3, 0.30984) +- (0, 0.1)
    (0.5, 0.13019) +- (0, 0.08)
    (0.56, 0.11655) +- (0, 0.08)
    (0.58, 0.10538) +- (0, 0.08)
    (0.7, 0.0781) +- (0, 0.08)
    (0.8, 0.05638) +- (0, 0.08)
    (0.9, 0.06051) +- (0, 0.08)
    (1.0, 0.04019) +- (0, 0.08)
    };
    \addlegendentry{Theoretical VQE N\'eel}

\addplot+[
    color=green,
    mark=square,
    smooth,
    error bars/.cd, y dir=both, y explicit
    ]
    coordinates {
    (0, 0.90)
    (0.2, 0.78) +- (0, 0.005934025160907104)
    (0.4, 0.555) +- (0, 0.006251360640749717)
    (0.5, 0.324) +- (0, 0.006875938204330133)
    (0.58, 0.09) +- (0, 0.007048439187476831)
    (0.8, 0.117) +- (0, 0.0072678028245958)
    (1.0, 0.03)
    };
    \addlegendentry{Experimental VQE N\'eel}

\addplot[
    color=babyblue,
    mark=circle,
    ]
    coordinates {
    (0,0.3687028481509969)(0.1,0.3578013343117531)(0.2,0.3421596380947529)(0.3,0.31818794851434007)(0.5,0.21353652520480468)(0.56,0.15936082933231205)(0.58,0.13887481715423206)(0.7,0.03139257557991701)(0.8,0.01081338640550112)(0.9,0.005605301477515712)(1.0,0.0035492787727874958)
    };
    \addlegendentry{Theoretical N\'eel}

\draw [babyblue] (axis cs:0.45,0.0) -- (axis cs:0.45,1.5);
\addlegendimage{line legend, color=babyblue}
\addlegendentry{Neel/Gapless Boundary}

\draw [red] (axis cs:0.56,0.0) -- (axis cs:0.56,1.5);
\addlegendimage{line legend, color=red}
\addlegendentry{Gapless/VBS}

\draw [black] (axis cs:0.61,0.0) -- (axis cs:0.61,1.5);
\addlegendimage{line legend, color=black}
\addlegendentry{VBS/collinear}

\end{axis}

\end{tikzpicture}
\caption{Comparison of Theoretical and Experimental N\'eel Order vs. different $J_{2}/J_{1}$ choices. Dashed vertical lines denote N\'eel/Gapless (babyblue), Gapless/VBS (red), and VBS/collinear (black) phase boundaries.}
\label{vqe_Neel}
\end{figure}
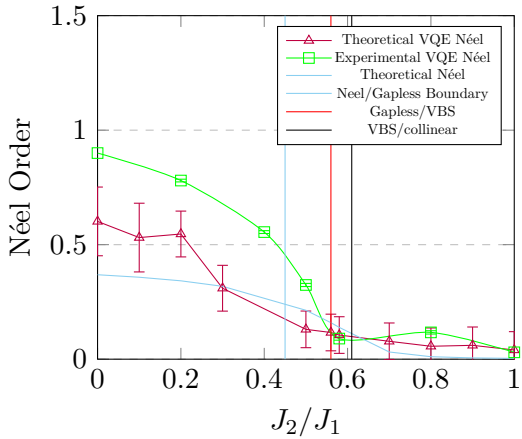

As a comparative baseline with theoretical predictions, one would expect the local and global correlation to intersect somewhere near the QSL phase at $J_{2}/J_{1} \approx 0.56$. This is due to the expectation that the global correlation should be lower relative to local correlation for $J_{2}/J_{1} < 0.56$ and greater for $J_{2}/J_{1} > 0.56$. In Figure \ref{vqe_correlation}, we observe such a behavior. One would also expect a maximum in the dimer order around the expected phase boundary of $J_{2}/J_{1} = 0.56$, as seen in Figure \ref{vqe_dimer}. Additionally, as we deviate from N\'eel ordering, we would anticipate a decrease in the N\'eel order parameter. The result in Figure \ref{vqe_Neel} agrees with these theoretical expectations. 

When compared with VQE, exact diagonalization has a more gradual and continuous transformation for the N\'eel order, the correlation functions, and the Dimer order. Perhaps, the rougher transitions are due to the variational nature of optimization VQE relies upon. When compared with averaged simulation performance, the IBM Rensselear system performs roughly on par with noiseless VQE simulation. The overall trend is captured by VQE simulation indicating the methods of error mitigation employed were successful. Note, in Figure \ref{vqe_correlation}, averaging of the VQE runs smears out the transition behavior near $J_{2}/J_{1} = 0.56$, illustrating the impact of VQE optimization error.

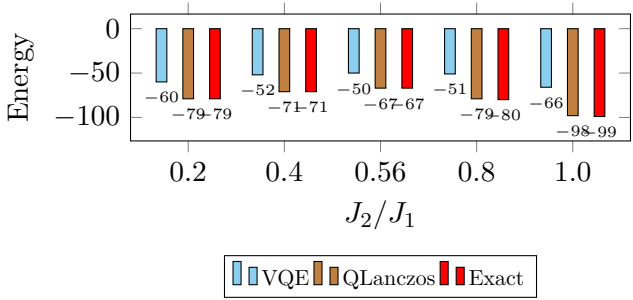
\begin{figure}
\begin{minipage}{0.48\textwidth}
\begin{tikzpicture}
\begin{axis}[
    ybar,
    enlargelimits=0.15,
    legend style={
        at={(0.5,-0.9)},
        anchor=north,legend columns=-1
    },
    ylabel={Energy},
    xlabel={$J_2 / J_1$},
    symbolic x coords={0.2,0.4,0.56,0.8,1.0},
    xtick=data,
    nodes near coords,
    nodes near coords align={vertical},
    nodes near coords style={font=\tiny},
    bar width=4pt,
    ymin=-110,
    ymax=0,
    width=\columnwidth, 
    height=0.4\columnwidth, 
    legend style={nodes={scale=0.75, transform shape}},
]

\addplot[fill=babyblue, bar shift=-10pt] coordinates {(0.2,-60) (0.4,-52) (0.56,-50) (0.8,-51) (1.0,-66)};

\addplot[fill=brown, bar shift=0pt] coordinates {(0.2, -79)(0.4, -71)(0.56, -67)(0.8, -79)(1.0, -98)};

\addplot[fill=red, bar shift=10pt] coordinates {(0.2,-79) (0.4,-71) (0.56,-67) (0.8,-80) (1.0,-99)};
\legend{VQE,QLanczos, Exact}

\end{axis}
\end{tikzpicture}
\caption{Energy values for VQE Simulation (as described in methods), QLanczos (15 Lanczos Vectors), and Exact Diagonalization (through Qiskit's NumPy Eigensolver) for varying $J_{2}/J_{1}$ choice.}
\label{vqe_energy_distribution}
\end{minipage}
\end{figure}

The frustrated phases of the square lattice near $J_{2}/J_{1} = 0.5$ tend to have higher energies than the other phases of the system. We will investigate if the expected trend is observed in VQE simulation. For the square lattice examined, we see that this trend is indeed observed, see Figure \ref{vqe_energy_distribution}. Rather interestingly, the collinear order seems to be the most inaccurate in terms of ground state approximation, according to the energy distribution in Figure \ref{vqe_energy_distribution}. Nevertheless, the expectation values of other observables do not significantly deviate from theoretical expectations. This indicates the VQE method's approximation of the ground state may represent the physics better than suggested purely by the energy expectation value.

In the results for the correlation values, dimer order, and N\'eel order, in Figures \ref{vqe_correlation}, \ref{vqe_dimer}, and \ref{vqe_Neel}, respectively, it is shown that the dimer order has the highest alignment with exact diagonalization and with VQE theory. This is followed by the correlation values and finally the N\'eel order. Before using dynamic decoupling and ZNE, the optimization results and observed expectation values experienced high variance throughout the $J_{2}/J_{1}$ region, with several values tending towards zero, indicative of noise effects. The observation showcases the dramatic improvement error mitigation can have. The greatest alignment between estimated and true ground state, occurs at the point of frustration, see Figure \ref{vqe_energy_distribution}. This indicates that noise does not necessarily relegate one to performing only well on phases well described by easily simulated AFM configurations. 
%
%

The current limitation rests on the VQE method itself and not the device's noise. In general, the VQE method has to play a balancing act between mitigating the intrinsic difficulties of quantum optimization and maintaining the representational power of the ansatz. These quantum optimization difficulties include barren plateaus and trapping swamp minima.

The Feulner and Hartmann ansatz on the periodic boundaries with 3 repetitions of the RX-CZ circular TwoLocal ansatz, was compared with our TwoLocal ansatz. Both approaches were limited to 500 optimization steps. With $J_{2}/J_{1}$ set to $0.5$, our TwoLocal ansatz yielded a $71.67\%$ approximation percentage relative to the true ground state found by exact diagonalization. The Feulner and Hartmann ansatz yielded a $60.00\%$ approximation percentage. The best performance was defined by trialing different repetitions amounts, 1 to 5, and different optimizers, including Nakanishi-Fujii-Todo (NFT) used in this study and Constrained Optimization By Linear Approximation optimizer (COBYLA) used in Feulner and Hartmann's study. Similar performance differences between the two ansatzes are found for open boundary variant of the Hamiltonian, tested by Feulner and Hartmann, at the same number of optimization steps. 

Additionally, as a sanity check, the ground state is solved, yielding the same expected theoretical ground state estimates of $-30.02$ K, illustrating consistency between this work and prior work for the employed Hamiltonian. \cite{feulner2022quantum} The performance matches the results of Feulner and Hartmann, within $<25\%$ after $3\times 10^{2}$ optimization steps. Compared to HFMA, the best prior state of the art, throughout the $J_{2}/J_{1}$ spectrum, we yield a $9.31\%$ advantage in ground state approximation.

This indicates consistency and raises an important point. While the ansatz tested may eventually converge to better solutions, practically, less theoretical ideal ansatz choices may be more feasible in the NISQ-era when optimization steps prove expensive. \cite{feulner2022quantum} Regardless, even without further efforts, it seems the trends are indeed captured with the VQE simulation results indicating the utility of the VQE algorithm for trend identification for the first time.

\begin{figure}[hbt]
\centering
\noindent\begin{tikzpicture}
\begin{axis}[
    xlabel={$J_{2}/J_{1}$},
    ylabel={N\'eel Order},
    xmin=0, xmax=1,
    ymin=0, ymax=0.4,
    legend pos=north east,
    grid style=dashed,
    ymajorgrids=true,
    legend style={nodes={scale=0.5, transform shape}}, 
]

\addplot[
    color=babyblue,
    mark=*,
    ]
    coordinates {
    (0,0.3687028481509969)(0.1,0.3578013343117531)(0.2,0.3421596380947529)(0.3,0.31818794851434007)
    (0.5,0.21353652520480468)(0.56,0.15936082933231205)(0.58,0.13887481715423206)(0.7,0.03139257557991701)
    (0.8,0.01081338640550112)(0.9,0.005605301477515712)(1.0,0.0035492787727874958)
    };
    \addlegendentry{Ground Truth}

\addplot[
    color=red,
    mark=square*,
    ]
    coordinates {
    (0.0, 0.31500616410132953)
    (0.1, 0.3077029206361398)
    (0.2, 0.29603131938714516)
    (0.30000000000000004, 0.27786602640203373)
    (0.4, 0.25076038362682185)
    (0.5, 0.21286660536769203)
    (0.6000000000000001, 0.16478101280923096)
    (0.7000000000000001, 0.1115228372339473)
    (0.8, 0.06215916667292638)
    (0.9, 0.025325258831741478)
    (1.0, 0.003990746430237547)
    };
    \addlegendentry{Adjusted N\'eel}

\end{axis}
\end{tikzpicture}
\caption{Adjusted, using the classical ZNE estimate correction mentioned in Methods,  N\'eel Order for different $J_{2}/J_{1}$ combinations. Ground Truth is determined through exact diagonalization.}
\label{vqe_corrected_N\'eel}
\end{figure}
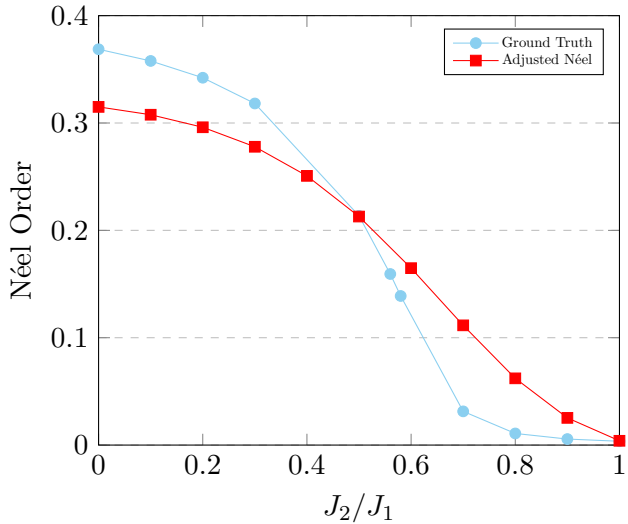

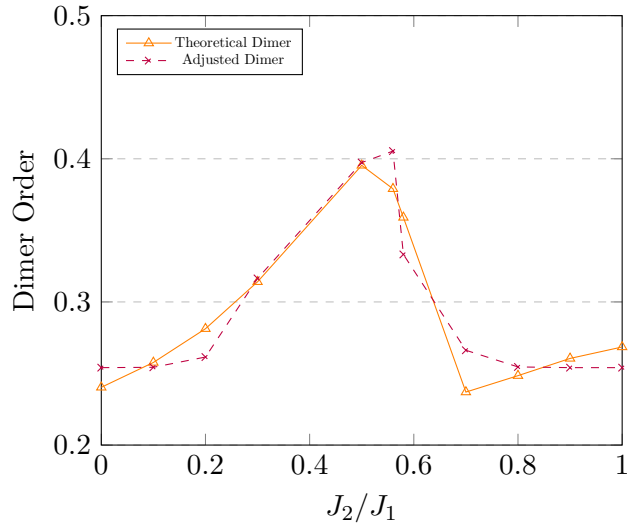
\begin{figure}[hbt]
\centering
\noindent\begin{tikzpicture}
\begin{axis}[
    xlabel={$J_{2}/J_{1}$},
    ylabel={Dimer Order},
    xmin=0, xmax=1,
    ymin=0.2, ymax=0.5,
    legend pos=north west,
    ymajorgrids=true,
    grid style=dashed,
    legend style={nodes={scale=0.5, transform shape}}, 
]

\addplot[color=orange, mark=triangle] coordinates {
    (0,0.24034282315311367) (0.1,0.2577057812956596) (0.2,0.2812798009644143) (0.3,0.3140678064223895) 
    (0.5,0.39541050358144636) (0.56,0.3790479338032299) (0.58,0.3590460781380115) (0.7,0.23702407576179274) 
    (0.8,0.24848603571885516) (0.9,0.26052818807036077) (1.0,0.26852262914756325)
};
\addlegendentry{Theoretical Dimer}

\addplot[color=purple, mark=x, dashed] coordinates {
    (0,0.25409084) (0.1,0.25433164) (0.2,0.26149016) (0.3,0.31642005) 
    (0.5,0.39734014) (0.56,0.4053922) (0.58,0.33318204) (0.7,0.26622943) 
    (0.8,0.25458506) (0.9,0.25409471) (1.0,0.25408829)
};
\addlegendentry{Adjusted Dimer}

\end{axis}
\end{tikzpicture}
\caption{Adjusted, using the classical ZNE estimate correction mentioned in Methods,  Dimer Order for different $J_{2}/J_{1}$ combination. Ground Truth is determined through exact diagonalization.}
\label{vqe_dimer_corrected}
\end{figure}

To combat error from hardware, not captured by ZNE error mitigation, experimental data from VQE is corrected using data from classical optimization. The need for correction becomes apparent when observing the correlation function in Figure \ref{vqe_correlation} going below $-1$ for $\frac{J_{2}}{J_{1}} < 0.4$. The observation is unphysical. In this case, expectation values at $J_{2}/J_{1} = 0$ and $J_{2}/J_{1} = 1$ can be easily computed. The reason being is that the classical state preparation of both the N\'eel and collinear states are known to overlap significantly with the true states. \cite{liu2020gapless} Exploiting this independence, the observables can be directly evaluated on the resulting state using classical optimization. The N\'eel order at $J_{2}/J_{1} = 0$ is thus estimated to be $0.351$, which is similar to the exact value of $0.369$. Likewise, the N\'eel order at $J_{2}/J_{1} = 1$ can be estimated to be $0$ comparing again nicely to the found value of $0.004$. Doing the same for the correlation values, assuming the endpoints can be efficiently estimated classically, allows the following corrections in Figures \ref{vqe_correlation_corrected}, \ref{vqe_corrected_N\'eel}, and \ref{vqe_dimer_corrected}.

\begin{figure}[hbt]
\centering
\noindent\begin{tikzpicture}
\begin{axis}[
    xlabel={$J_{2}/J_{1}$},
    ylabel={Correlation Function},
    xmin=0, xmax=1,
    ymin=-1, ymax=0.5,
    legend pos=north west,
    ymajorgrids=true,
    grid style=dashed,
    scale=0.9,
    legend style={nodes={scale=0.5, transform shape}}, 
]

\addplot[color=babyblue, mark=o] coordinates {
    (0,-0.7017802005268026) (0.1,-0.700895102070927) (0.2,-0.6971621329415036) (0.3,-0.6874489287103996)
    (0.5,-0.6053182704365818) (0.56,-0.5283801649185245) (0.58,-0.4884064509773419) (0.7,-0.16847269275029914)
    (0.8,-0.08564425664057519) (0.9,-0.059269946065780885) (1.0,-0.04639692549529759)
};
\addlegendentry{Classical Local Z}

\addplot[color=red, mark=o] coordinates {
    (0,-0.6048370236562555) (0.1,-0.607037685841411) (0.2,-0.6079479627385287) (0.3,-0.6056843140065034)
    (0.5,-0.5593346760799276) (0.56,-0.5042545724970893) (0.58,-0.47371994933616957) (0.7,-0.2154575517425132)
    (0.8,-0.14657844173225357) (0.9,-0.12421851626803389) (1.0,-0.11315498398359748)
};
\addlegendentry{Classical Global Z}

\addplot[color=green, mark=x, dashed] coordinates {
    (0,-0.7574360809574855) (0.1,-0.79710388480747) (0.2,-0.8823982765085902) (0.3,-1.000901787161984)
    (0.4,-0.9979150213661924) (0.5,-0.7217971768809276) (0.6,-0.36985904993767016) (0.7,-0.1476242314420998)
    (0.8,-0.046265600885595504) (0.9,-0.0059488499008620195) (1.0,0.009257528189359628)
};
\addlegendentry{Adjusted VQE Local}

\addplot[color=cyan, mark=x, dashed] coordinates {
    (0,-0.6430606172031778) (0.1,-0.6795056002246168) (0.2,-0.7580548698394668) (0.3,-0.874332259237191)
    (0.4,-0.9071475985414618) (0.5,-0.7103216351958627) (0.6,-0.41978517814336325) (0.7,-0.22249881841653096)
    (0.8,-0.1284013504292319) (0.9,-0.08982576570995293) (1.0,-0.07493383442176091)
};
\addlegendentry{Adjusted VQE Global}

\end{axis}
\end{tikzpicture}
\caption{Adjusted, using the classical ZNE estimate correction mentioned in Methods, Local and Global Z Correlation Order for different $J_{2}/J_{1}$ combinations. Ground Truth is determined through exact diagonalization.}
\label{vqe_correlation_corrected}
\end{figure}
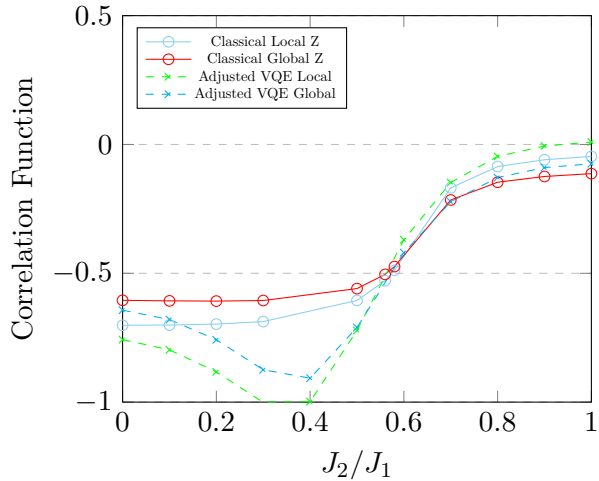

Overall, in the case where our experiments are corrected by classically-reinforced ZNE, Figure \ref{vqe_correlation_corrected}, \ref{vqe_corrected_N\'eel}, and \ref{vqe_dimer_corrected}, one can see a dramatic improvement in the ability to capture the quantitative properties of the given correlation functions, while still retaining qualitative agreement. We demonstrate with this novel method that post-processing with classical expectation values may be able to enhance the simulation veracity of near-term devices.

\begin{figure}
\begin{minipage}{0.5\textwidth}
\begin{tikzpicture}
\begin{axis}[
    xlabel={$J_{2}/J_{1}$},
    ylabel={Local Z Correlation Value},
    xmin=0, xmax=1,
    ymin=-1, ymax=0.5,
    legend pos=north west,
    ymajorgrids=true,
    grid style=dashed,
    scale=0.9,
    legend style={nodes={scale=0.7, transform shape}}, 
]

\addplot[color=babyblue, mark=o] coordinates {
    (0,-0.506) (0.2,-0.43) (0.4,-0.316) (0.5,-0.265)(0.56,-0.265) (0.8,-0.15) (1,-0.1237) (0.8,-0.15)
};
\addlegendentry{Local Z Real Evolution 15 Dim}

\addplot[color=red, mark=square*] coordinates {
    (0,-0.6924) (0.1,-0.705777) (0.2,-0.6830) (0.4,-0.69857) (0.5,-0.5578) (0.56,-0.49244) 
    (0.58,-0.45894) (0.7,-0.189) (0.8,-0.115) (0.9,-0.0718) (1.0,-0.0525)
};
\addlegendentry{Local Lanczo 15 Dim}

\addplot[color=purple, mark=triangle] coordinates {
    (0,-0.702) (0.1,-0.700) (0.2,-0.697) (0.4,-0.687) (0.5,-0.605) (0.56,-0.528) 
    (0.58,-0.488) (0.7,-0.168) (0.8,-0.0856) (0.9,-0.059) (1.0,-0.046)
};
\addlegendentry{Theoretical Local}

\end{axis}
\end{tikzpicture}
\end{minipage}
\begin{minipage}{0.5\textwidth}
\begin{tikzpicture}
\begin{axis}[
    xlabel={$J_{2}/J_{1}$},
    ylabel={Global Z Correlation Value},
    xmin=0, xmax=1,
    ymin=-1, ymax=0.5,
    legend pos=north west,
    ymajorgrids=true,
    grid style=dashed,
    scale=0.9,
    legend style={nodes={scale=0.7, transform shape}}, 
]

\addplot[color=red, mark=o] coordinates {
    (0,-0.5979) (0.1,-0.61) (0.2,-0.595) (0.4,-0.6111) (0.5,-0.5229) (0.56,-0.476) 
    (0.58,-0.4495) (0.7,-0.23024) (0.8,-0.169)
};
\addlegendentry{Global Lanczo 15 Dim}

\addplot[color=babyblue, mark=square*] coordinates {
    (0,-0.439) (0.2,-0.371) (0.4,-0.2966) (0.5,-0.258) (0.56,-0.259) (0.8,-0.167)(1,-0.153) 
};
\addlegendentry{Global Z Real Evolution 15 Dim}

\addplot[color=purple, mark=triangle] coordinates {
    (0,-0.6048) (0.1,-0.607) (0.2,-0.608) (0.4,-0.559) (0.5,-0.504) (0.56,-0.473) 
    (0.58,-0.215) (0.7,-0.147) (0.8,-0.123)(1.0,-0.11315498398359748)
};
\addlegendentry{Theoretical Global}

\end{axis}
\end{tikzpicture}
\caption{Local Z Value (Top) and Global Z Value (Bottom) for different $J_{2}/J_{1}$ combinations for the real evolution basis and Lanczos basis.}
\label{local_z_krylov}
\label{global_z_krylov}
\end{minipage}
\end{figure}

\begin{figure}
\centering
\noindent\begin{tikzpicture}
\begin{axis}[
    xlabel={$J_{2}/J_{1}$},
    ylabel={Neel Order},
    xmin=0, xmax=1,
    ymin=0, ymax=0.5,
    legend pos=north east,
    ymajorgrids=true,
    grid style=dashed,
    scale=0.9,
    legend style={nodes={scale=0.7, transform shape}}, 
]

\addplot[color=red, mark=o] coordinates {
    (0,0.363) (0.1,0.362) (0.2,0.336) (0.4,0.330) (0.5,0.192266) (0.56,0.1482) 
    (0.58,0.131) (0.7,0.041) (0.8,0.022959) (0.9,0.01102895) (1.0,0.00920589)
};
\addlegendentry{Neel Lanczo 15 Dim}

\addplot[color=babyblue, mark=square*] coordinates {
    (0,0.2698) (0.2,0.22327) (0.4,0.13263) (0.5,0.10309) (0.56,0.0986) (0.8,0.059)(1,0.04647) 
};
\addlegendentry{Neel Real Evolution 15 Dim}

\addplot[color=purple, mark=triangle] coordinates {
    (0,0.3687) (0.1,0.3578) (0.2,0.3421) (0.3,0.3135) (0.5,0.159) (0.56,0.139) 
    (0.58,0.0314) (0.7,0.0108) (0.8,0.0056) (0.9,0.0035) (1.0,0.0035492787727874958)
};
\addlegendentry{Theoretical Neel}

\end{axis}
\end{tikzpicture}
\caption{Neel Order for different $J_{2}/J_{1}$ combinations for the Lanczos basis versus the real time evolution basis.}
\label{neel_krylov}
\end{figure}
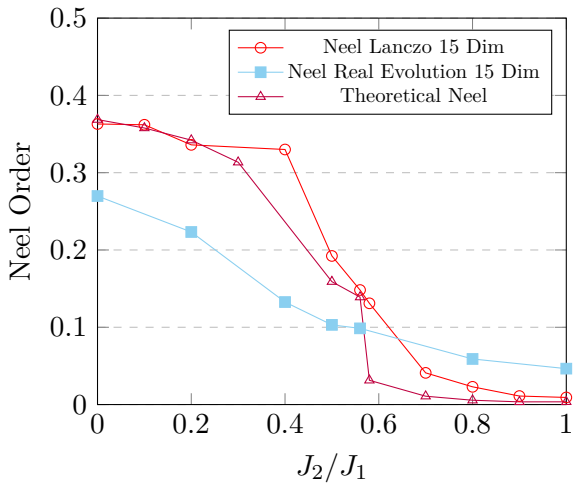

\begin{figure}
\centering
\noindent\begin{tikzpicture}
\begin{axis}[
    xlabel={$J_{2}/J_{1}$},
    ylabel={Dimer Order},
    xmin=0, xmax=1,
    ymin=0.2, ymax=0.5,
    legend pos=north west,
    ymajorgrids=true,
    grid style=dashed,
    scale=0.9,
    legend style={nodes={scale=0.7, transform shape}}, 
]

\addplot[color=red, mark=o] coordinates {
    (0,0.237) (0.1,0.25779) (0.2,0.27678) (0.4,0.309) (0.5,0.38596) (0.56,0.3428) 
    (0.58,0.3422) (0.7,0.22477) (0.8,0.2348) (0.9,0.25459) (1.0,0.263)
};
\addlegendentry{Dimer Lancoz 15 Dim}

\addplot[color=babyblue, mark=square*] coordinates {
    (0,0.204) (0.2,0.2139) (0.4,0.2317) (0.5,0.2327) (0.56,0.2364)  (0.8,0.224)(1,0.231) 
};
\addlegendentry{Dimer Real Evolution 15 Dim}

\addplot[color=purple, mark=triangle] coordinates {
    (0,0.24) (0.1,0.2577) (0.2,0.28127) (0.3,0.314) (0.5,0.395) (0.56,0.379) 
    (0.58,0.359) (0.7,0.237) (0.8,0.248) (0.9,0.2605) (1.0,0.268522)
};
\addlegendentry{Theoretical Dimer}

\end{axis}
\end{tikzpicture}
\caption{Dimer Order for different $J_{2}/J_{1}$ combinations for the Lanczos basis versus the real time evolution basis.}
\label{dimer_order}
\label{dimer_krylov}
\end{figure}
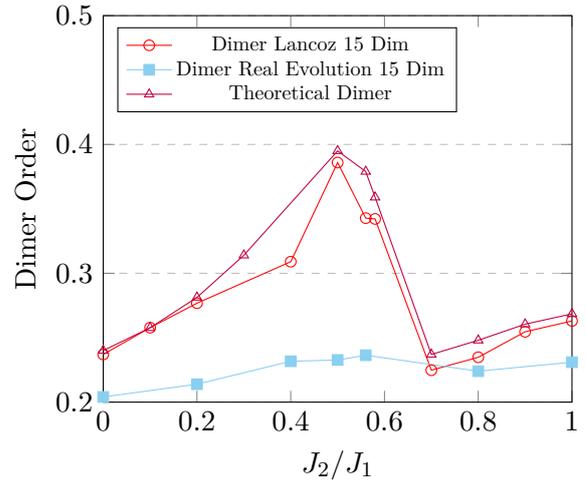

\subsection{QLanczos}

\begin{table}[htb]
\centering
\label{table:fidelities}
\begin{tabular}{|c|c|}
\hline
$k$ & Fidelity \\ \hline
1 & 0.647 \\ \hline
2 & 0.969 \\ \hline
3 & 0.489 \\ \hline
4 & 0.743 \\ \hline
5 & 0.507 \\ \hline
6 & 0.925 \\ \hline
\end{tabular}
\caption{Fidelity versus $H^k$ using Lanczos vectors using Variational Fast Forwarding (VFF) in Basis Reconstruction, Eq.~ 13 cost value: 0.336, after 558 iteration of the NFT optimizer.}
\label{lanczos_approximation}
\end{table}

As Figure \ref{vqe_energy_distribution} shows, there exists theoretical promise in using the Lanczos basis in its ability to capture the energy observable. The finding is consistent with Kirby, Motta, and Mezzacapo's work, which showed a similar quick convergence of their method for the $J_1-J_2$ system. \cite{kirby2023quantum} The capture of the energy observable is significant in its own respect, but a different pragmatic question of the ability for the method to capture other physical observables of interest naturally follows. To examine this, we compare the Real Evolution Krylov basis expectations, the Lanczos Krylov basis expectations, and the theoretical expectations, see Figures \ref{local_z_krylov}, \ref{global_z_krylov}, \ref{neel_krylov}, and \ref{dimer_krylov}. Overall, it seems the Real Time Evolution, although performing similarly in predicting energy, within $5\%$, performs poorly in capturing other observables. This is apparent when increasing the number of Krylov vectors. Additionally, the alignment in Figure \ref{local_z_krylov} (bottom) seems to be the most divergent, particular near frustration, indicating that possibly local observables converge quicker than global observables near frustration points.

To explore simulating with the Quantum Lanczos algorithm, the question of implementation on IBM's system needs to be addressed. Direct implementation of the $\mathbf{U}$, $\mathbf{R}$, and $\mathbf{G}$ operator construction proposed by Kirby, Motta, and Mezzacapo on hardware requires a depth of approximately 1483 with around 375 two-qubit gates for $k = 1$ for a Hamiltonian $\mathbf{H}$ with only XX, XZ, ZX, ZZ terms. This makes it beyond the realm of feasibility on the IBM Rensselaer device for our system of interest. 

However, a significant portion of this depth and two-qubit requirement is unavoidable in following the paper's proposed scaling behavior of $\approx O(kn^2)$ for increased system size is observed. The choice was also made to minimize the number of qubits involved in the Variational Fast Forwarding (VFF) optimization at the cost of increased classical computation time. 
 
To make the implementation more practical on IBM's system, the variational fast-forwarding approach (VFF) described above is utilized producing the results in Table \ref{lanczos_approximation}. Note, fidelity in Table \ref{lanczos_approximation} references the average fidelity of the Krylov vectors defined at the end of Variational Fast Forwarding, section 1.3.6. We find that even $k$ tends to yield better approximations even when performing different optimization runs. The better performance at even $k$ seems to indicate that even repetitions are simply easier to model for the proposed $\textbf{RU}$ operator. Note, the parity observation only applies to the Variational Fast-Forwarding result. 

An important point made in the work by Filip, Ramo, and Fitzpatrick, which becomes notable when examining the fidelities. \cite{filip2024variational} The inaccuracies made in replicating $\textbf{H}^{k}$ do not necessarily translate to poor ability to approximate the ground state unlike the case of VQE. \cite{filip2024variational}  The reason is that one is still likely to maintain a portion of the action of $\textbf{H}^{k}$ allowing the construction of an approximate Lanczos basis, which may still be able to serve as a support subspace. \cite{filip2024variational}
 
Direct noiseless quantum simulation of the method gives $-22.47$ K, comparing favorably with the ground truth of $-24.00$ K within three Lanczos vectors. With Linear Extrapolation, 10000 shots, and a Qiskit transpiler optimization level of 3, the circuit depth of the compiled circuit with the VFF approach was less than $700$. The approximate solution after $3$ vectors was found to be $-21.80$ K, suggesting the method remains advantageous in spite of the less than ideal fidelity created by VFF. As a comparison, a $60\%$ fidelity direct implementation of similar depth using approximate quantum compilation yielded $-4.26$ K, suggesting that error due to the circuit depth persists as a problem for naive method implementation. To mitigate the loss of positive definiteness through error or numerical precision issues, $\textbf{S}$ is replaced with $\textbf{S}' = \textbf{S} \pm \epsilon\textbf{I}$ for some small $\epsilon$, theoretically $10^{-3}$. Experimentally the derived $\textbf{S}$ required an $\epsilon$ of $1.1$. This works by shifting the eigenvalues by exactly $\epsilon$ to offset the positive definiteness loss. 

With QLanczos, converging more rapidly than real-time evolution, see Figures \ref{local_z_krylov}, \ref{global_z_krylov}, \ref{neel_krylov}, and \ref{dimer_krylov}, an interesting question arises: how does one best optimize between theoretical convergence performance to desirable resulting observables? \cite{yeter2021,Tkachenko_2024,10.1007/978-981-15-9927-9_27,PRXQuantum.2.010333,Nandy:2024htc} The issue is striking because Real Time Evolution provided similar energy expectations yet produced dramatically different performance in generalizing to other expectation values. Practically speaking, real-time evolution has greater appeal, as it does not require block encoding and is an area of research more widely investigated. 
Its poor ability to capture other physical observables besides energy also raises the question: is the energy observable too naive of a metric to gauge the near-term practicality of such Krylov basis choices?

It seems natural to consider both the generality of the algorithm for other systems and the scaling of its performance for larger system sizes. The requirement for $\epsilon$-addition to maintain positive definiteness of $\textbf{S}$ raises the question on how to best stabilize the Quantum Lanczos approach in the face of real-world error if one wishes to use the approach in the NISQ framework. And, although QLanczos is the most theoretically promising out of the tested methods, a general implementation faces too much upfront cost and the experimental approach used here to mitigate the issue is limited in scale to small systems. But, the recent work of Kriby, Motta, and and Mezzacpo gives hope to the idea that such limitations may be mitigated with better implementation making the issue more of a practical than theoretical limitation of the method. Like in the case of VQE, it is also conceivable that system-specific features may be exploitable to further reduce hardware requirements.

\section{Conclusion and Outlook}

\paragraph{} In this study, we have tackled the problem of simulating frustrated spin systems on near-term quantum devices. By employing a modified version of VQE with a classical warm-start approach, we aimed to address the challenges posed by high entanglement and diminishing spectral gaps characteristic of these spin systems intrinsic to classical methods.

Our results demonstrate for the first time that, despite the inherent noise present in current NISQ-era hardware, it is possible to capture the significant trends of the $J_{1}-J_{2}$ Heisenberg model, particularly in the identification of phase transitions. The success of the warm-starting optimization, as evidenced by the consistency of energy evaluations and correlation functions, reinforces the viability of this approach. It also suggests that even approximate classical assumptions can provide a substantial advantage in guiding quantum optimization routines towards more accurate solutions. Additionally, our original technique for ZNE correction, called Classically-Reinforced ZNE, has shown the benefit of utilizing known or easily generated expectation values for phases to greatly improve quantum simulations. This creates avenues for making present day NISQ devices viable for materials simulation by combining their strengths to offset their weaknesses. 

We demonstrate that while noise adversely affects the VQE outcomes, its mitigation through techniques like zero-noise extrapolation and dynamic decoupling is effective. Our experiments suggest that further refinements in error mitigation could enable even more accurate simulations on quantum hardware. 

Comparisons between exact diagonalization and our VQE results reveal that while VQE may not replicate exact results, it faithfully reproduces the physically significant trends and qualitative results, such as the phase transitions. These findings underscore the potential of VQE to serve as a practical tool for studying complex quantum phenomena. Also, when comparing with the results for the Quantum Computed Moments VQE correction scheme, it becomes clear that the introduced correction schema is a more resource-efficient way of integrating existing physical knowledge to improving the outputs produced by VQE (see Appendix section B). 

The QLanczos method was found to be the most theoretically and experimentally promising, although issues with its implementation's scaling present barriers to near-term practicality. The need for VFF to query the target function directly also implicitly limits the scale to current simulation procedure. Although, it would be interesting if the optimization objective for VFF could be optimized more cheaply, i.e. optimize for convergence on a certain Clifford circuit family. Regardless of the practical challenges, the strong alignment with physically-relevant observables, quick theoretical convergence assurances, and desirable theoretical scaling strengthens the credence of the method.

In the future, due to the success of the VFF method, it may be of interest to move to less direct variational methods like variational quantum imaginary time evolution (Var-QITE) and variational quantum phase estimation (Var-QPE). These methods may be better able to approximate the physical system by solving a related but not identical, and possibly more tractable, quantum-classical optimization problem to efficiently project into a classically-solvable subspace. \cite{zhou2017quantum, fedorov2022vqe}

In conclusion, our work demonstrates the practicality of employing NISQ quantum algorithms for simulating complex quantum systems on current quantum hardware. In particular, it provides a basis for the study of exotic spin systems. It is hoped that the refinements proposed here, suggestions on future works, along with better hardware, will pave the way for increasingly accurate simulations for trend analysis on NISQ devices.

\section{Acknowledgements}

\paragraph{} This material is based upon work supported by the NSF CAREER award under grant number 2044842. The authors thank Haimeng Zhang, Darren Kwee, and Meltem Tolunay for their assistance in utilizing the IBM Rensselear device, developing the algorithmic approach, and discussing research trajectory.

\section{Data Availability}

\paragraph{} Template code for the experimental and theoretical data collection is available at: \textit{https://github.com/dylansheils/Spin-1-2-J1-J2-Quantum-Simulation}. Raw data was generated by IBM cloud using "IBM Rensselear". Derived data from IBM cloud supporting the findings of this study are available from the corresponding author D.S. upon reasonable request.

\printbibliography

\appendix

\section{Warm-Starting Technical Details}

\paragraph{} The independent subspaces would be described by a Kronecker sum of 2D subspaces as opposed to a Kronecker product of 2D subspaces. Thus, an approximation would be assuming, for any position $k$, $S_{k}$ is a 2D complex valued vector representing the independent qubit state. With the dimensionality reduction, classical optimization gains feasibility. The question of how to perform optimization over $S_{k}$ for each site $k$ natural arises. 

\paragraph{} Recognizing each qubit as being associated with a position on a Bloch sphere allows the following parameterization of each spin on each site: $\ket{\psi}_{k}(\theta_{k}, \phi_{k}) = \cos{\theta_{k}}\ket{0} + \sin{\theta_{k}}e^{i\phi_{k}}\ket{1}$. This reduces the problem to minimizing the sum of each sites energy with the constraints $0 \le \phi_{k} \le \pi$ and $0 \le \theta_{k} \le 2\pi$ for all sites $k$. As each qubit has a $\phi_{k}$ and $\theta_{k}$ parameter, the problem reduces to a continuous, bounded $2Q$ variable quadratic program.

\section{Variational Fast-Forwarding Technical Details}

\paragraph{} Taking from the variational form suggested by Filip, Ramo, and Fitzpatrick \cite{filip2024variational}: 

\begin{align}
  \label{variational_form}
    \textbf{A}^{k} = \textbf{W}(\overrightarrow{\theta}, \overrightarrow{\phi}) \textbf{D}(k\overrightarrow{\gamma}) \textbf{W}^{\dagger}(\overrightarrow{\theta}, \overrightarrow{\phi})
\end{align}

\paragraph{} Originally, the formulation of this variational ansatz assumed one was decomposing the real time evolution operator associated with time-step $n$, $e^{iHn\Delta t}$, with uniform time samplings of size $\Delta t$. \cite{filip2024variational} In this case, $\textbf{W}$, describes pairwise interactions of $\phi$ and $\theta$ defining rotations similar to $U(\theta_{k}, \phi_{k})$, with a diagonal component $\textbf{D}$, $Z$ rotations and $ZZ$ pairwise coupling interactions, decomposition. For our purposes, the circuit preparation for $\ket{\psi_{0}}$ is absorbed into this construction. To optimize this circuit, we follow the introduced cost function cited by Filip, Ramo, and Fitzpatrick \cite{filip2024variational}:

\begin{align}
  \label{optimization_objective}
    f(\overrightarrow{\theta}, \overrightarrow{\gamma}) = 1 - \frac{1}{n}\sum_{k = 1}^{n} |\bra{\psi_{0}}(\textbf{A}^{k})^{\dagger}\textbf{T}_{k}(\textbf{H})\ket{\psi_{0}}|^{2}
\end{align}

\paragraph{} In this situation, $|\bra{\psi_{0}}(\textbf{A}^{k})^{\dagger}\textbf{T}_{k}(\textbf{H})\ket{\psi_{0}}|^{2}$ represents the fidelity between the desired output vector and ansatz output vector. We compute the expectation value by using a Hadamard test for the real and imaginary components separately then combining the results afterwards.

\section{Quantum Lanczos Technical Details}

\subsection{Conceptual Overview}

\paragraph{} The desired matrix element computations needed for the Quantum Lanczos algorithm described earlier is the following \cite{kirby2023quantum}:

\begin{align}
  \label{krylov}
    \bra{\psi_{0}} \textbf{T}_{k}(H) \ket{\psi_{0}} = \begin{cases}
        \left<\textbf{R}\right>_{k/2} & \text{, $k$ is even} \\
        \left<\textbf{U}\right>_{k/2} & \text{, $k$ is odd}
    \end{cases}
\end{align}

\paragraph{} In this case, $\ket{\psi_{k/2}} = (RU)^{\lfloor k/2 \rfloor}(\ket{G}_{a} \bigotimes \ket{\psi}_{0})$ and $\left<\textbf{O}\right>_{k/2} = \bra{\psi_{k/2}} \textbf{O} \ket{\psi_{k/2}}$ for observable $\textbf{O}$. $\textbf{R} = (2\ket{G}_{a}\bra{G}_{a} - \textbf{I}_{a}) \bigotimes \textbf{I}_{s}$ where $\textbf{I}_{a}$ represents the identity operator operating on the ancillary system and $\textbf{I}_{s}$ represents the identity operation on the qubits representing the system and the state $\ket{G} = \sum_{i = 0}^{T - 1} \sqrt{\alpha_{i}} \ket{i}_{a}$. $i$ denotes the binary encoding of the index of Pauli term $i$. The matrix elements are determined as follows: 

\begin{align}
  \label{S_krylov}
    \textbf{S}_{ij} = \frac{1}{2}(\left<\textbf{T}_{i + j}(H)\right>_{0} + \left<\textbf{T}_{|i - j|}(H)\right>_{0})
\end{align}

\begin{align}
  \label{H_krylov}
    \textbf{H}_{ij} = \frac{1}{4}\left( \left<\textbf{T}_{i + j + 1}(H)\right>_{0} + \left<\textbf{T}_{|i + j - 1|}(H)\right>_{0} \right. \nonumber \\
    \left. + \left<\textbf{T}_{|i - j + 1|}(H)\right>_{0} + \left<\textbf{T}_{|i - j - 1|}(H)\right>_{0} \right)
\end{align}

\subsubsection{Creation of Block Encoding for $\textbf{H}$}

\paragraph{} A problem arises when implementing the QLanczos method naively on hardware—the possible non-unitary nature of $\textbf{H}$. \cite{camps2022fable} The general formulation of the encoding is the following:

\begin{align}
  \label{block_encoding}
    \textbf{U} = \begin{bmatrix} \textbf{H} & * \\
    * & *
    \end{bmatrix}
\end{align}

The encoding represents the fact that the non-unitary operator $\textbf{H}$ can be transformed into a higher dimensional unitary transformation, $\textbf{U}$, with $a$ ancillary qubits. However, the application of $\textbf{H}$ only succeeds when the ancillary measurement yields the state $\ket{0}_{a}$. Note, $*$ denotes required entries, specific to the choice of $\textbf{H}$, to ensure norm-preservation of $\textbf{U}$. Transformations with such embedding schemes unifies many different quantum computing algorithms within the framework of quantum singular value transformation (QSVT) and form the basis for quantum signal processing (QSP). \cite{camps2022fable} A condition for the construction of a valid $\textbf{U}$ is that: $||\textbf{A}||_{2} \le 1$. To ensure this is met, Kirby, Motta, and Mezzacapo impose the constraint that $\sum_{i=0}^{T-1} |\alpha_{i}| = \sum_{i = 0}^{T-1} \alpha_{i} = 1$ through $\alpha_{i} = \frac{\beta_{i}}{\sum_{i=0}^{T-1} |\beta_{i}|}$ where $\beta_{i}$ are the inital coefficients associated with the Pauli string deconstruction of $\textbf{H} = \sum_{i = 0}^{T - 1}\beta_{i}P_{i}$. \cite{kirby2023quantum} 

\subsection{Creation of $\textbf{G}$}

\paragraph{} The definition of mapping $\ket{0}$ to $\ket{G}$ is used to define a single column of $\mathbf{G}$, while the rest of the columns are defined using the Householder transformation. QR factorization is used to ensure the unitary property is preserved for the choice of $\mathbf{G}$ in the construction. 

\section{Quantum Computed Moments}

\subsection{Algorithmic Explanation}

\paragraph{} To test other correction schemes to VQE, the method of quantum computed moments was attempted. Proposed by Vallury, Jones, Hill, \textit{et. al}, the Quantum Computed Moments algorithm exploits the infimum theorem found in Lanczos cumulant expansion analysis to generate better estimates of energy and of observables. \cite{Vallury2020} It produces such estimates using a finite number of Hamiltonian moment expectations, $\left<\textbf{H}^{k}\right>$, which are used to generate a finite Lanczos cumulant expansion. The following gives the cumulants from the Hamiltonian moment expectations:

\begin{align}
  \label{cumulants}
    c_{n} = \left<\textbf{H}^{n}\right> - \sum_{p = 0}^{n - 2} \binom{n - 1}{p}c_{p + 1}\left<\textbf{H}^{n - 1 - p}\right>
\end{align}

 From these cumulants, one can estimate the ground state energy using:

\begin{align}
  \label{ground_state}
    E_{0} \approx c_{1} - \frac{c_{2}^{2}}{c_{3}^{2} - c_{2}c_{4}}(\sqrt{3c_{3}^{2} - 2c_{2}c_{4}} - c_{3})
\end{align}

 To measure other observables, say $\textbf{O}$, one perform the following trick:

\begin{align}
  \label{observable}
    \textbf{H}_{\lambda} = \textbf{H} + \lambda\textbf{O}
\end{align}

 By taking a finite difference estimation with $\lambda = \epsilon$ using the observable $\textbf{H}_{\lambda}$, one can approximate the target observable, $\textbf{O}$, with the corrected ground state estimation, according to the work done by Vallury and Hollenberg. \cite{Vallury2023}

\subsection{Performance Analysis}

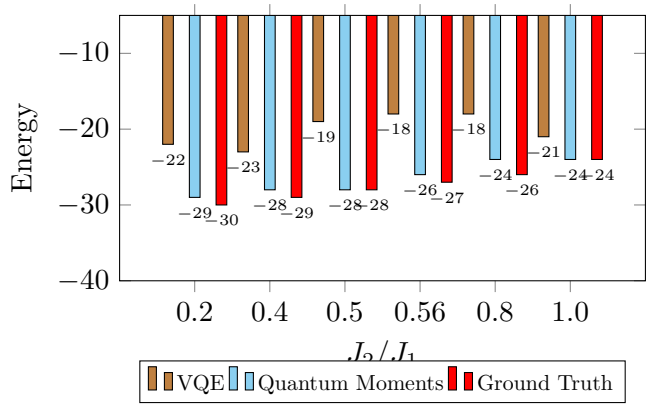
\begin{figure}
\centering
\begin{minipage}{0.5\textwidth}
\begin{tikzpicture}
\begin{axis}[
    ybar,
    enlargelimits=0.2,
    legend style={
        at={(0.5,-0.3)},
        anchor=north,legend columns=-1
    },
    ylabel={Energy},
    xlabel={$J_2 / J_1$},
    symbolic x coords={0.2, 0.4, 0.5, 0.56, 0.8, 1.0},
    xtick=data,
    nodes near coords,
    nodes near coords align={vertical},
    nodes near coords style={font=\tiny},
    bar width=4pt,
    ymin=-35,
    ymax=-10,
    width=\columnwidth,
    height=0.6\columnwidth,
    legend style={nodes={scale=0.75, transform shape}},
]

\addplot[fill=brown, bar shift=-10pt] coordinates {(0.2,-22) (0.4,-23) (0.5,-19) (0.56,-18) (0.8,-18) (1.0,-21)};
\addplot[fill=babyblue, bar shift=0pt] coordinates { (0.2,-29) (0.4,-28) (0.5,-28) (0.56,-26) (0.8,-24) (1.0,-24)};
\addplot[fill=red, bar shift=10pt] coordinates {(0.2,-30) (0.4,-29) (0.5,-28) (0.56,-27) (0.8,-26) (1.0,-24)};
\legend{VQE, Quantum Moments, Ground Truth}

\end{axis}
\end{tikzpicture}
\caption{Energy distribution for varying $J_{2}/J_{1}$ for a 3x3 lattice using TwoLocal (RX, CZ, 3 repetitions), comparing VQE, Quantum Moments Enhanced VQE, and Ground Truth simulation.}
\label{vqe_moments_energy_distribution}
\end{minipage}
\end{figure}

\paragraph{} Although creation of a Lanczos subspace is an enticing proposal, the block encoding scheme proves to be somewhat restrictive to the method. However, the sidestepping of the variational optimization problem seems to be of interest. To improve the variational approach, one can exploit Hamiltonian moments, using those moments to compute Lanczos cumulant expansion terms, and then use a finite order truncation of such terms to generate a better estimate to the ground state, using the VQE algorithm as a trial ansatz generator. \cite{Vallury2020} The method is referred to as Quantum Computed Moments. \cite{Vallury2020} Additionally, exploiting the redefinition of $\textbf{H}$ to $\textbf{H} + \lambda\textbf{O}$ for a desirable observable $\textbf{O}$ and then taking an finite difference estimate with respect to $\lambda$ allows one to generate an estimate with respect to arbitrary observables as well. \cite{Vallury2023}

To explore the performance of the algorithm within a computational study, the energy distribution figure of Figure \ref{vqe_energy_distribution} was regenerated for the 3x3 lattice, Figure \ref{vqe_moments_energy_distribution}. It is seen in Figure \ref{vqe_moments_energy_distribution} that significant improvements in ground state estimation occur with a order 4 Lanczos cumulant expansion. The improvement is quite dramatic, reducing the maximum error throughout the domain of $J_{2}/J_{1}$, from $26.66\%$ to $3.5\%$. Vallery, Jones, and \textit{et. al}. suggested, to reduce the naively exponential growth of Pauli terms with $\textbf{H}^{k}$'s enumeration a conversion of the Hamiltonian into a Tensor Product Basis form which one finds commuting measurement groups and forms circuit estimates from within each clique. \cite{Vallury2020} Although promising and theoretically efficient, employing the schema using the clique cover of commuting Tensor Product Basis representation terms for the $J_{2}/J_{1}$ basis in the case of $N \ge 16$ requires running the estimates of around $O(1000)$ circuits with an average of $O(100)$ Pauli terms. 

To mitigate these challenges, our approach attempts a modified form of importance estimation inspired by Chakraborty's single-ancillary observable estimation schema. \cite{Chakraborty:2023vtr} In this case, the observable is expressed as a Pauli terms with sampling occurring based on their associated coefficient in the linear combination relative to the 1-norm of all coefficients. In much the same way, our modified form of Quantum Moments uses only the first few dominate cliques and samples for a fixed number of Pauli strings based on importance sampling thereby improving the efficiency of Vallury's proposal. \cite{Vallury2020} 

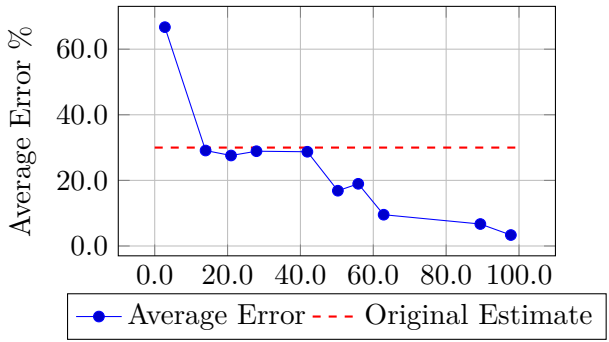
\begin{figure}
\centering
\begin{tikzpicture}
\begin{axis}[
    width=0.9\columnwidth,
    height=0.6\columnwidth,
    xlabel={\% Pauli Terms},
    ylabel={Average Error \%},
    grid=major,
    legend style={
        at={(0.5,-0.15)},
        anchor=north,legend columns=-1
    },
    x tick label style={/pgf/number format/.cd,%
        fixed, fixed zerofill, precision=1},
    y tick label style={/pgf/number format/.cd,%
        fixed, fixed zerofill, precision=1},
]

\addplot coordinates {
    (2.793296089385475, 66.71187946636682)
    (13.966480446927374, 29.101219178650595)
    (20.94972067039106, 27.585456996559248)
    (27.932960893854748, 28.894022625481036)
    (41.89944134078212, 28.72782265844759)
    (50.27932960893855, 16.848162324243955)
    (55.865921787709496, 18.957400857287357)
    (62.849162011173185, 9.535060339926632)
    (89.3854748603352, 6.698287493857347)
    (97.76536312849163, 3.3405610030528)
};

\addplot[thick, red, dashed] coordinates {(0,30) (100,30)};

\legend{Average Error, Original Estimate}

\end{axis}

\end{tikzpicture}
\caption{Average Error of Ground State Prediction vs. \% Pauli Terms}
\label{fig:error_vs_pauli}
\end{figure}

To gauge the dependency of the percentage of Pauli terms versus the percent error in ground state estimation, plotted in Figure \ref{fig:error_vs_pauli}, it becomes clear that around 50\% is sufficient in the current situation to significantly improve the ground state estimation. However, it seems that under-sampling less than 5\% of the Pauli terms actually worsens the estimate. Naively, the growth per moment is: 108, 2755, 18127, 32896 for $k = 1, 2, 3, 4$. This requires, in IBM's system, 103, 368, 638, 720 circuit estimates, respectively. However, an advantage of the current approach is that, due to the constraints that $3c_{3}^{2} - 2c_{2}c_{4} \geq 0$ and $E_{0}(c_{1},c_{2},c_{3},c_{4}) \leq E_{0}(c_{1})$, one can actually discard experimental values that violate such constraints and identify cases of under-sampling. Performing binary search on the error rate allows one to efficiently trade-off computation time with estimation improvement, with only a logarithmic runtime overhead.

When running a TwoLocal optimization with 3 repetitions using CX and RZ gates for 600 optimization steps with the COBYLA optimizer for a $J_{2}/J_{1} = 0.5$ 9-site instance, it was found that, before the Quantum Moments algorithm was used the energy estimate was $-19.10$ K when experimentally ran but $-21.04$ K theoretically. Running the above approach generated an estimate of $-23.11$ K experimentally and $-27.74$ K theoretically greatly reducing the error in both theory and in practice for ground state estimation. For reference, ground truth energy was $-28.65$ K found through diagonalization. 

To gauge the effective change in the energy expectation, the average integral dot product between the adjusted and ground truth expectation values was taken. For global Z and local Z, the averaged functional 1-norm inner products were $0.363$ and $0.293$, respectively. And, for the N\'eel and Dimer order, averaged inner products of $0.051$ and $0.090$ were found, respectively. Averaging gives $0.199$ with a variance of $0.023$ for these observables. A rough estimate then would be $19.9\% \pm 2.3\%$ error in ground state approximation. The majority of error comes from Figure \ref{vqe_correlation_corrected}. When acknowledging the poor performance near $0 \le \frac{J_{2}}{J_{1}} \le 0.6$ likely resulted from the unphysical result in Figure \ref{vqe_correlation} near the domain, defining an modified error without this outlier, gives an estimate of $7\% \pm 1\%$. Comparing with the percent error from Figure \ref{vqe_moments_energy_distribution}, one finds $3\% \pm 1\%$. 

\subsection{Quantum Moments versus Classically-Reinforced VQE}

\paragraph{} Overall, the Quantum Moments algorithm has a greater affinity in improving the energy estimates from VQE in theory when compared with Classically-Reinforced VQE, but it comes at a significant runtime cost which makes it impractical for larger Hamiltonians in the near-term at large system sizes. Beyond the runtime cost, in practice, the error in estimating these Hamiltonian moments scales with the Pauli terms. This leads to a decay in realizable advantage from $3.2\%$ theoretically to $16.2\%$ experimentally. Although the theoretical performance is better than expected from Classically-Reinforced VQE, Classically-Reinforced VQE realizes the advantage with an error of $7\% \pm 1\%$.

\end{document}